
\input harvmac

\Title{\vbox{\baselineskip12pt
\hbox{QMW-PH-98-37}
\hbox{CAMS/98-04}
\hbox{hep-th/9808069}}}
{\vbox{\centerline{ Branes, Times and Dualities}}}

\baselineskip=12pt
\centerline {Chris M. Hull$^1$\footnote{$^a$}{C.M.Hull@qmw.ac.uk}
 and Ramzi R. Khuri$^{1,2}$\footnote{$^b$}{R.R.Khuri@qmw.ac.uk.
Address as of September 1st, 1998: Department of Natural Sciences,
Baruch College, 17 Lexington Ave., New York, NY, 10010, USA.}}
\medskip
\centerline{\sl $^1$Department of Physics}
\centerline{\sl Queen Mary and Westfield College}
\centerline{\sl Mile End Road}
\centerline{\sl London E1 4NS UK}
\medskip
\centerline{\sl $^2$Center for Advanced Mathematical Sciences}
\centerline{\sl American University of Beirut}
\centerline{\sl Beirut, Lebanon}

\bigskip
\centerline{\bf Abstract}
\medskip
\baselineskip = 20pt

Dualities link M-theory, the 10+1 dimensional strong coupling limit of the
IIA string, to other 11-dimensional theories in signatures 9+2 and 6+5,
and to type II string theories in all 10-dimensional signatures. We study
the Freund-Rubin-type compactifications and brane-type solutions of these
theories, and find that branes with various world-volume signatures are
possible. For
example, the 9+2 dimensional M* theory has membrane-type solutions with
world-volumes of signature (3,0) and (1,2), and a solitonic solution  with
world-volume signature (5,1).

\Date{August 1998}

\font\mybb=msbm10 at 10pt
\def\bbbb#1{\hbox{\mybb#1}}
\def\Z {\bbbb{Z}}
\def\R {\bbbb{R}}

\def\ie{{\it i.e.,}\ }

\def\({\left (}
\def\){\right )}
\def\[{\left [}
\def\]{\right ]}

\def \aa {\alpha}
\def \bb {\beta}

\def \ss {\sigma}
\def \tt {\tau}

\def \ti {\tilde}

\def \2 {{1 \over 2}}
\def \3 {{1 \over 3}}
\def \4 {{1 \over 4}}
\def \5 {{1 \over 5}}
\def \6 {{1 \over 6}}
\def \7 {{1 \over 7}}
\def \8 {{1 \over 8}}
\def \9 {{1 \over 9}}
\def \0 { \infty}

\def\++ {{(+)}}
\def \- {{(-)}}
\def\+-{{(\pm)}}

\def\ek {\eqn\abc}

\def \pa {\partial}

\def \qq {\qquad}

\def\unit{\hbox to 3.3pt{\hskip1.3pt \vrule height 7pt width .4pt \hskip.7pt
\vrule height 7.85pt width .4pt \kern-2.4pt
\hrulefill \kern-3pt
\raise 4pt\hbox{\char'40}}}

\def\nup#1({Nucl.\ Phys.\  {\bf B#1}\ (}

\def\ie{{\it i.e.,}\ }

\def\hphi{\hat{\phi}}
\def\ha{\hat{a}}
\def\hg{\hat{\gamma}}

\def\td{\tilde{d}}
\def\tF{\tilde{F}}
\def\tH{\tilde{H}}

\lref\chamblin{A. Chamblin and R. Emparan, Phys. Rev. {\bf D55} (1997) 754.}

\lref\emparan{R. Emparan, Nucl. Phys. {\bf B490} (1997) 365.}

\lref\guven{R. Guven, Phys. Lett. {\bf B276} (1992) 49.}

\lref\bko{C.M. Hull, Phys. Lett. {\bf B39} (1984) 139;
E. A. Bergshoeff, R. Kallosh and T. Ortin,
Phys. Rev. {\bf D47} (1993) 5444.}

\lref\bgppr{J. Barrett, G. W. Gibbons, M. J. Perry, C. N. Pope
and P. Ruback, Int. J. Mod. Phys. {\bf A9} (1994) 1457.}

\lref\eleven{M. J. Duff, P. S. Howe, T. Inami and K. Stelle, Phys. Lett.
{\bf B191} (1987) 70.}

\lref\stain{H. Lu, C. N. Pope, E. Sezgin, and K. S. Stelle,
Nucl. Phys. {\bf B456} (1995) 669; see also H. Lu, C. N. Pope,
and K. S. Stelle, Nucl. Phys. {\bf B481} (1996) 313 and references
therein.}

\lref\ddduff{M. J. Duff and J. X. Lu, Nucl. Phys. {\bf B416} (1994) 301.}

\lref\chrisone{C. M. Hull, hep-th/9806146.}

\lref\christwo{C. M. Hull, hep-th/9807127.}

\lref\mal{J. Maldacena,   hep-th/9711200.}

\lref\nextpaper{C. M. Hull and R. R. Khuri, in preparation.}


\lref\rust{R. R. Khuri and R. C. Myers, Nucl. Phys. {\bf B466} (1996) 60.}

\lref\GSW {M. B. Green, J. H. Schwarz and E. Witten,
{\it Superstring Theory}, Cambridge University Press, Cambridge (1987).}

\lref\prep{See M. J. Duff, R. R. Khuri and J. X. Lu, Phys. Rep.
{\bf B259} (1995) 213 and references therein.}

\lref\PKT{P.K. Townsend, Phys. Lett. {\bf B350}  (1995) 184.}

\lref\moore{G. Moore, hep-th/9305139,9308052.}

\lref\CJ{E. Cremmer and B. Julia, Phys. Lett. {\bf 80B} (1978) 48; Nucl.
Phys. {\bf B159} (1979) 141.}

\lref\julia{B. Julia in {\it Supergravity and Superspace}, S.W. Hawking
and M. Ro$\check c$ek, C.U.P.
Cambridge,  (1981). }

\lref\julec{B.  Julia, hep-th/9805083.}

\lref\HT{C.M. Hull and P.K. Townsend, hep-th/9410167.}

\lref\gibrap{G.W. Gibbons, hep-th/9803206.}

\lref\buscher{T. H. Buscher, Phys. Lett. {\bf 159B} (1985) 127,
Phys. Lett. {\bf B194}
 (1987), 51 ; Phys. Lett. {\bf B201}
 (1988), 466.}

\lref\rocver {M. Ro\v cek and E. Verlinde, Nucl. Phys.
{\bf B373} (1992), 630.}

 \lref\givroc {A. Giveon, M.
Ro\v cek, Nucl. Phys. {\bf B380} (1992), 128.}

\lref\alv{E. Alvarez, L. Alvarez-Gaum\' e,
J.L. Barbon and Y. Lozano,  Nucl. Phys. {\bf B415} (1994)
71.}

\lref\TD {A. Giveon, M. Porrati and E. Rabinovici, Phys. Rep. {\bf 244}
(1994) 77.}

\lref\HJ{C. M. Hull  and B. Julia, hep-th/9803239.}

\lref\CPS{E. Cremmer,  I.V. Lavrinenko,  H. Lu,
C.N. Pope,  K.S. Stelle and  T.A. Tran, hep-th/9803259.}

\lref\Stelle{
K. S. Stelle, hep-th/9803116.}

\lref\ddua {J. Dai, R.G. Leigh and J. Polchinski, Mod. Phys. Lett. {\bf
A4} (1989) 2073.}

\lref\dsei{ M. Dine, P. Huet and N. Seiberg, Nucl. Phys. {\bf B322}
(1989) 301.}

\lref\gibras{  G.W. Gibbons and D.A. Rasheed, hep-th/904177.}

\lref\Bob{B.S. Acharya, M. O'Loughlin and B. Spence, Nucl.
Phys. {\bf B503} (1997) 657; B.S. Acharya, J.M. Figueroa-O'Farrill, M.
O'Loughlin and B. Spence,
hep-th/9707118.}

\lref\thom{M. Blau and G. Thompson, Phys. Lett. {\bf B415} (1997) 242.}

\lref\hawtim{S.W. Hawking,  Phys.Rev. {\bf D46 } (1992) 603.}

\lref\dinst{M.B.  Green, Phys. Lett. {\bf  B354}
(1995) 271,  hep-th/9504108;
M.B.~Green and M.~Gutperle,   Phys.Lett. B398(1997)69, hep-th/9612127;
M.B.~Green and M.~Gutperle,   hep-th/9701093,
G.~Moore, N.~Nekrasov and  S.~Shatahvilli,   hep-th/9803265;
E. Bergshoeff and  K. Behrndt, hep-th/9803090.}

\lref\sevbrane{G.W. Gibbons, M.B.  Green and M.J. Perry, Phys.Lett.
B370 (1996) 37, hep-th/9511080.}

\lref\beck{K.~Becker, M.~Becker and  A.~Strominger, hep-th/9507158,
Nucl.Phys. {\bf 456} (1995) 130.}

\lref\dinstcal{K.~Becker, M.~Becker, D.R.~Morrison, H.~Ooguri, Y.~Oz and
Z.~Yin,
  hep-th/9608116, Nucl.Phys. {\bf 480} (1996) 225; H.~Ooguri and C.~Vafa,
hep-th/9608079, Phys. Rev. Letts. 77(1996) 3296; M.~Gutperle,
hep-th/9712156.}

\lref\adsstab{P. Breitenlohner and D.Z. Freedman, Phys. Lett. {\bf 115B}
(1982) 197; Ann. Phys. {\bf 144} (1982) 197; G.W. Gibbons, C.M. Hull and
N.P. Warner, Nucl. Phys. {\bf B218} (1983) 173.}

\lref\mal{J. Maldacena,   hep-th/9711200.}

\lref\OS{K. Osterwalder and R. Schrader, Phys. Rev. Lett.
{\bf 29} (1972) 1423; Helv. Phys. Acta{\bf 46}
 (1973) 277;
CMP {\bf 31} (1973) 83 and CMP {\bf 42} (1975) 281.
K. Osterwalder,  in G. Velo and A. Wightman (Eds.)
Constructive Field Theory - Erice lectures 1973,
Springer-Verlag Berlin 1973; K. Osterwalder in {\it Advances in Dynamical
Systems and
Quantum Physics}, Capri conference, World Scientific 1993. }

\lref\vanwick{P. van Nieuwenhuizen and A. Waldron, Phys.Lett. B389 (1996)
29-36, hep-th/9608174. }

\lref\KT{T. Kugo and P.K. Townsend, Nucl. Phys. {\bf B221} (1983) 357.}

\lref\yam{J. Yamron, Phys. Lett. {\bf B213} (1988) 325.}

\lref\vafwit{C. Vafa and E. Witten, Nucl. Phys. {\bf B431} (1994) 3-77,
hep-th/9408074.}

\lref\sing{ L. Baulieu, I. Kanno and I. Singer, hep-th/9704167.}

\lref\Seib{N. Seiberg,
hep-th/9705117.}%

\lref\vans{K. Pilch, P. van Nieuwenhuizen and M. Sohnius, Commun. Math. Phys.
{\bf 98} (1985) 105.}

\lref\luk{J. Lukierski and A. Nowicki,   Phys.Lett. {\bf 151B}  (1985)  382.}

\lref\witt{E. Witten, hep-th/9802150.}

\lref\bergort{E. Bergshoeff, C.M. Hull and T. Ortin, Nucl. Phys. {\bf B451}
(1995) 547, hep-th/9504081.}

\lref\huto{C.M. Hull, Nucl. Phys. {\bf B509} (1998) 252, hep-th/9702067. }

\lref\asp{P. Aspinwall, Nucl. Phys. Proc. Suppl. {\bf  46}  (1996) 30,
hep-th/9508154; J. H. Schwarz, hep-th/9508143.}

\lref\fvaf{C. Vafa, Nucl. Phys. {\bf 469} (1996) 403.}

\lref\GravDu{C.M. Hull, Nucl. Phys. {\bf B509} (1997) 252, hep-th/9705162.}
\lref\bergnin{ E.~Bergshoeff and J.P.~van der Schaar,
                {  hep-th/9806069}.}

\lref\ythe{C.M. Hull, Nucl.Phys. {\bf B468}  (1996) 113  hep-th/9512181;
A. A. Tseytlin Nucl.Phys.
{\bf B469}  (1996) 51  hep-th/9602064.}

\lref\mythe{I. Bars,  Phys. Rev. {\bf D54} (1996) 5203, hep-th/9604139;
hep-th/9604200;
Phys.Rev. {\bf D55} (1997) 2373 hep-th/9607112 .}

\lref\huku{C.M. Hull and R.R. Khuri, in preparation.}

\lref\inter{G. W. Gibbons and P. K. Townsend, Phys. Rev. Lett.
{\bf 71} (1993) 3754; M. J. Duff, G. W. Gibbons and P. K. Townsend,
Phys. Lett. {\bf B332} (1994) 321; G. W. Gibbons, G. T. Horowitz and
P. K. Townsend, Class. Quan. Grav. {\bf 12} (1995) 297.}

\lref\adssol{M. Gunaydin and N. Marcus, Class. Quant. Grav {\bf 2} (1985) L11;
 H.J. Kim, L.J. Romans
and P. van Nieuwenhuizen, Phys. Rev. {\bf D32} (1985) 389.}

\lref\CW{C.M. Hull and N. P. Warner, Class. Quant. Grav. {\bf 5} (1988) 1517.}

\lref\poly{S.S. Gubser,  I. R. Klebanov and  A. M. Polyakov, hep-th/9802109.}

\lref\hor{ G. Horowitz and A. Strominger, Nucl. Phys {\bf B360} (1991) 197.}

\lref\huten{C.M. Hull, Phys. Lett. {\bf B357 } (1995) 545, hep-th/9506194.}

\lref\gibhaw{G.W. Gibbons and S.W. Hawking, Phys. Rev. {\bf D15} (1977) 2738.}

\lref\kall{P. Claus, R. Kallosh, J. Kumar, P.K. Townsend and A. van Proeyen,
hep-th/9801206.}

\lref\witkk{E. Witten, Nucl. Phys. {\bf B195} (1982) 481.}

\lref\haweuc{S.W. Hawking, in {\it General Relativity}, ed. by S.W. Hawking
and W. Israel, Cambridge University Press, 1979.}

\lref\dufstel{M.J. Duff and K.S. Stelle, Phys. Lett. {\bf B253} (1991) 113.}

\lref\horwit{P. Horava and E. Witten, Nucl. Phys. {\bf B460} (1996) 506.}

\lref\wit{E. Witten, Nucl. Phys. {\bf B443} (1995) 85.}

\lref\blencowe{M. P. Blencowe and M. J. Duff, Nucl. Phys.
{\bf B10} (1988) 387.}

\lref\fr{P. G. O. Freund and M. A. Rubin, Phys. Lett. {\bf B97} (1980) 233;
F. Englert,  Phys. Lett. {\bf B119} (1982) 339; see also M. J. Duff,
B. E. W. Nilsson and C. N. Pope, Phys. Rep. {\bf 130} vols. 1 \& 2
(1986) 1 and references therein.}

\lref\witads{E. Witten, hep-th/9802150.}

\lref\pvt{K. Pilch, P. van Nieuwenhuizen and P. K. Townsend,
Nucl. Phys. {\bf B242} (1984) 377.}

\lref\cjs{E. Cremmer, B. Julia and J. Scherk,
Phys. Lett.{\bf B76} (1978) 409.}

\lref\gps{D. J. Gross and M. J. Perry, Nucl. Phys. {\bf B226} (1983) 29;
Phys. Rev. Lett. {\bf 51} (1983) 87.}


%
%
%
%
\newhelp\stablestylehelp{You must choose a style between 0 and 3.}%
\newhelp\stablelinehelp{You should not use special hrules when stretching
a table.}%
\newhelp\stablesmultiplehelp{You have tried to place an S-Table inside another
S-Table.  I would recommend not going on.}%
%
%
\newdimen\stablesthinline
\stablesthinline=0.4pt
\newdimen\stablesthickline
\stablesthickline=1pt
%
%
\newif\ifstablesborderthin
\stablesborderthinfalse
\newif\ifstablesinternalthin
\stablesinternalthintrue
\newif\ifstablesomit
\newif\ifstablemode
\newif\ifstablesright
\stablesrightfalse
%
%
\newdimen\stablesbaselineskip
\newdimen\stableslineskip
\newdimen\stableslineskiplimit
%
%
\newcount\stablesmode
\newcount\stableslines
\newcount\stablestemp
\stablestemp=3
\newcount\stablescount
\stablescount=0
\newcount\stableslinet
\stableslinet=0
%
%
%
\newcount\stablestyle
\stablestyle=0
%
%
\def\stablesleft{\quad\hfil}%
\def\stablesright{\hfil\quad}%
%
%
\catcode`\|=\active%
%
%
\newcount\stablestrutsize
\newbox\stablestrutbox
\setbox\stablestrutbox=\hbox{\vrule height10pt depth5pt width0pt}
\def\stablestrut{\relax\ifmmode%
                         \copy\stablestrutbox%
                       \else%
                         \unhcopy\stablestrutbox%
                       \fi}%
%
%
\newdimen\stablesborderwidth
\newdimen\stablesinternalwidth
\newdimen\stablesdummy
\newcount\stablesdummyc
\newif\ifstablesin
\stablesinfalse
%
%
\def\begintable{\stablestart%
  \stablemodetrue%
  \stablesadj%
  \halign%
  \stablesdef}%
\def\stablesadj{%
  \ifcase\stablestyle%
    \hbox to \hsize\bgroup\hss\vbox\bgroup%
  \or%
    \hbox to \hsize\bgroup\vbox\bgroup%
  \or%
    \hbox to \hsize\bgroup\hss\vbox\bgroup%
  \or%
    \hbox\bgroup\vbox\bgroup%
  \else%
    \errhelp=\stablestylehelp%
    \errmessage{Invalid style selected, using default}%
    \hbox to \hsize\bgroup\hss\vbox\bgroup%
  \fi}%
\def\stablesend{\egroup%
  \ifcase\stablestyle%
    \hss\egroup%
  \or%
    \hss\egroup%
  \or%
    \egroup%
  \or%
    \egroup%
  \else%
    \hss\egroup%
  \fi}%
\def\stablestart{%
  \ifstablesin%
    \errhelp=\stablesmultiplehelp%
    \errmessage{An S-Table cannot be placed within an S-Table!}%
  \fi
  \global\stablesintrue%
  \global\advance\stablescount by 1%
  \message{<S-Tables Generating Table \number\stablescount}%
  \begingroup%
  \stablestrutsize=\ht\stablestrutbox%
  \advance\stablestrutsize by \dp\stablestrutbox%
  \ifstablesborderthin%
    \stablesborderwidth=\stablesthinline%
  \else%
    \stablesborderwidth=\stablesthickline%
  \fi%
  \ifstablesinternalthin%
    \stablesinternalwidth=\stablesthinline%
  \else%
    \stablesinternalwidth=\stablesthickline%
  \fi%
  \tabskip=0pt%
  \stablesbaselineskip=\baselineskip%
  \stableslineskip=\lineskip%
  \stableslineskiplimit=\lineskiplimit%
  \offinterlineskip%
  \def\borderrule{\vrule width \stablesborderwidth}%
  \def\internalrule{\vrule width \stablesinternalwidth}%
  \def\thinline{\noalign{\hrule height \stablesthinline}}%
  \def\thickline{\noalign{\hrule height \stablesthickline}}%
  \def\trule{\omit\leaders\hrule height \stablesthinline\hfill}%
  \def\ttrule{\omit\leaders\hrule height \stablesthickline\hfill}%
  \def\tttrule##1{\omit\leaders\hrule height ##1\hfill}%
  \def\stablesel{&\omit\global\stablesmode=0%
    \global\advance\stableslines by 1\borderrule\hfil\cr}%
  \def\el{\stablesel&}%
  \def\elt{\stablesel\thinline&}%
  \def\eltt{\stablesel\thickline&}%
  \def\elttt##1{\stablesel\noalign{\hrule height ##1}&}%
  \def\elspec{&\omit\hfil\borderrule\cr\omit\borderrule&%
              \ifstablemode%
              \else%
                \errhelp=\stablelinehelp%
                \errmessage{Special ruling will not display properly}%
              \fi}%
  \def\stmultispan##1{\mscount=##1 \loop\ifnum\mscount>3 \stspan\repeat}%
  \def\stspan{\span\omit \advance\mscount by -1}%
  \def\multicolumn##1{\omit\multiply\stablestemp by ##1%
     \stmultispan{\stablestemp}%
     \advance\stablesmode by ##1%
     \advance\stablesmode by -1%
     \stablestemp=3}%
  \def\multirow##1{\stablesdummyc=##1\parindent=0pt\setbox0\hbox\bgroup%
    \aftergroup\emultirow\let\temp=}
  \def\emultirow{\setbox1\vbox to\stablesdummyc\stablestrutsize%
    {\hsize\wd0\vfil\box0\vfil}%
    \ht1=\ht\stablestrutbox%
    \dp1=\dp\stablestrutbox%
    \box1}%
  \def\stpar##1{\vtop\bgroup\hsize ##1%
     \baselineskip=\stablesbaselineskip%
     \lineskip=\stableslineskip%
     \lineskiplimit=\stableslineskiplimit\bgroup\aftergroup\estpar\let\temp=}%
  \def\estpar{\vskip 6pt\egroup}%
  \def\stparrow##1##2{\stablesdummy=##2%
     \setbox0=\vtop to ##1\stablestrutsize\bgroup%
     \hsize\stablesdummy%
     \baselineskip=\stablesbaselineskip%
     \lineskip=\stableslineskip%
     \lineskiplimit=\stableslineskiplimit%
     \bgroup\vfil\aftergroup\estparrow%
     \let\temp=}%
  \def\estparrow{\vfil\egroup%
     \ht0=\ht\stablestrutbox%
     \dp0=\dp\stablestrutbox%
     \wd0=\stablesdummy%
     \box0}%
  \def|{\global\advance\stablesmode by 1&&&}%
  \def\|{\global\advance\stablesmode by 1&\omit\vrule width 0pt%
         \hfil&&}%
  \def\vt{\global\advance\stablesmode by 1&\omit\vrule width \stablesthinline%
          \hfil&&}%
  \def\vtt{\global\advance\stablesmode by 1&\omit\vrule width
\stablesthickline%
          \hfil&&}%
  \def\vttt##1{\global\advance\stablesmode by 1&\omit\vrule width ##1%
          \hfil&&}%
  \def\vtr{\global\advance\stablesmode by 1&\omit\hfil\vrule width%
           \stablesthinline&&}%
  \def\vttr{\global\advance\stablesmode by 1&\omit\hfil\vrule width%
            \stablesthickline&&}%
  \def\vtttr##1{\global\advance\stablesmode by 1&\omit\hfil\vrule width ##1&&}%
  \stableslines=0%
  \stablesomitfalse}
\def\stablesdef{\bgroup\stablestrut\borderrule##\tabskip=0pt plus 1fil%
  &\stablesleft##\stablesright%
  &##\ifstablesright\hfill\fi\internalrule\ifstablesright\else\hfill\fi%
  \tabskip 0pt&&##\hfil\tabskip=0pt plus 1fil%
  &\stablesleft##\stablesright%
  &##\ifstablesright\hfill\fi\internalrule\ifstablesright\else\hfill\fi%
  \tabskip=0pt\cr%
  \ifstablesborderthin%
    \thinline%
  \else%
    \thickline%
  \fi&%
}%
\def\endtable{\advance\stableslines by 1\advance\stablesmode by 1%
   \message{- Rows: \number\stableslines, Columns:  \number\stablesmode>}%
   \stablesel%
   \ifstablesborderthin%
     \thinline%
   \else%
     \thickline%
   \fi%
   \egroup\stablesend%
\endgroup%
\global\stablesinfalse}
%
%



\newsec{Introduction}

The strong coupling limit of
the type IIA superstring theory is a theory in 10+1 dimensions that we
will refer to as M-theory.
The various superstring theories can be obtained from this by
compactifying and going to the
boundaries of the moduli space, giving a geometrical interpretation of many of
the string
theory dualities {\HT}. For example, the IIA string theory arises from
compactifying
M-theory on a circle and taking the zero radius limit {\wit},
the
IIB string theory arises from compactifying
M-theory on a 2-torus and taking the zero size limit \asp\ and the
$E_8\times E_8$ heterotic string
arises from compactifying
M-theory on $S^1/\Z_2$ and taking the zero radius limit \horwit.
 In \refs{\chrisone,\christwo},
this was generalised to the case of M-theory compactified on tori
with Lorentzian signature.
Surprisingly, going to   boundaries of the moduli space in which
Lorentzian cycles shrink to
zero size leads to new theories, in many of which the space-time
signature has more than one time. These include  versions of M-theory
in signatures 9+2 and 6+5 and
type II string theories in various 10-dimensional spacetime signatures.
All are linked by an intricate web of dualities and all emerge as different
limits of the same
underlying theory.
The set of theories   can be thought of as different
real forms
of a single complexified theory, and each  has a supergravity theory with 32
supersymmetries
arising
as
a field
theory limit. In this way, maximal supergravity theories in 10 or 11
dimensions with
various signatures are obtained, and each can be studied in its own right.
Each of the  forms of M-theory or type II string theories have brane
solutions of various
world-volume signatures, and our purpose here is to find and investigate
these solutions. We will restrict ourselves here to those solutions that
preserve 16 of the 32
supersymmetries, which are the analogues of the BPS states that play a key
role in the Lorentzian
signature theories. We will verify that the dualities linking the various
theories also relate the
brane solutions, giving a useful consistency check on the duality
structure unravelled in
\refs{\chrisone,\christwo},
and giving insights into the structure of these theories.

In \refs{\chrisone, \christwo},
timelike compactifications of M-theory and type II string
theories on Lorentzian
tori $T^{n,1}$ (with $n$ spatial circles  and one timelike circle) were
considered, extending the work  in  \refs{\HJ,\CPS} on timelike
compactifications of  supergravities. If,
as was assumed, such classical solutions with periodic time are consistent
backgrounds for
M-theory or
string theory, then the limits in which various cycles degenerate can be
considered.
The type IIA theory on a timelike circle gives, in the limit in which
the circle shrinks
to zero size, a T-dual type IIB* string theory, while the timelike
T-dual of the IIB
theory is a IIA* theory.
The IIA* and IIB* theories are formulated in 9+1 spacetime dimensions,
and have
supergravity limits similar to those of the IIA,B supergravities, but with
the signs of
some of the  kinetic terms reversed; in particular, the RR fields have
 kinetic terms of
the wrong sign.
The IIB* theory can be obtained from M-theory compactified on $T^{1,1}$
 in the limit
in which the Lorentzian 2-torus shrinks to zero size
\refs{\chrisone,\christwo}.
If M-theory is compactified on a 3-torus $T^{2,1}$, then the limit in which
the 3-torus
shrinks to zero size gives an M* theory in 9+2 dimensions. This same theory
 arises as
the strong coupling limit of the IIA* string theory, in which an extra
 time dimension
opens up \christwo,
in a way that is analogous to the way an extra space dimension opens
 up in the
strong coupling limit of type IIA theory \wit.
Similarly, starting from the M* theory in 9+2 dimensions and compactifying on
 a Euclidean
3-torus $T^3$ gives, in the limit in which the 3-torus shrinks to zero size,
an M$'$  theory
in 6+5 dimensions \christwo.
Similar steps lead to M-type theories in signatures 5+6, 2+9 and 1+10 also,
 but these are
equivalent to the theories with the reversed signature.

Starting from the M-theories in signatures 10+1, 9+2 and 6+5, and
compactifying on either a
spacelike or a timelike circle and taking the limit in which the circle
 shrinks to zero
size gives a set of type IIA-like string theories in signatures
10+0, 9+1, 8+2, 6+4 and 5+5 (together with the reversed forms).
There are two forms of the 9+1 dimensional theory,
 the IIA and IIA* theories,
and although there is a similar pair of forms of the theory in 5+5 dimensions,
they are related by
a field redefinition (including the signature-reversing $g_{\mu \nu } \to -
g_{\mu \nu } $)
and so are equivalent.
Compactifying the M-theories on 2-tori of various signatures
 and taking the limit
 in which both radii
tend  to zero size gives IIB-like theories in signatures 9+1, 7+3
and 5+5.
In signatures 9+1 and 5+5, there is a IIB theory and a IIB* theory,
together with a
IIB$'$ theory which is the strong coupling limit of the IIB* theory in
 either case.
Thus there is a set of string theories in 10 dimensions with various
signatures that are
related by timelike T-dualities and S-dualities, and this list gives all
such theories that
can be obtained in this way.

These theories have branes of various world-volume signatures, and they can
be formally
obtained from
the usual branes of M-theory or type II theory by dualities. For example,
 the timelike
T-duality takes the D-branes of the type II theories, with Lorentzian
world-volume
signature, to the
E-branes of the type II* theories, which have Euclidean world-volume
signatures.
We will refer to a brane with world-volume signature $(s,t)$ ($s$ positive
signature space dimensions and $t$ negative
signature timelike ones)
occuring in a theory with spacetime
 signature $(S,T)$, with $S\ge s,T\ge t$, as an $(s,t)$-brane.  Some branes in
non-standard signatures
were discussed in \blencowe, but most of the solutions that will be
discussed here were beyond the scope of \blencowe.

 There are also solutions which are products of generalised de Sitter spaces of
various signatures,
all of which are coset spaces $SO(p,q)/SO(p-1,q)$ with the $SO(p,q)$-invariant
metric; when these
have two sheets, we take one connected component. These include $d$-dimensional
de Sitter space
\eqn\eka{dS_d={SO(d,1) \over SO(d-1,1)}{ } ,}
$d$-dimensional anti-de Sitter space
\eqn\ekb{AdS_d={SO(d-1,2) \over SO(d-1,1)}{ },}
the
$d$-sphere
\eqn\ekc{S^d={SO(d+1) \over SO(d)}{ },}
the $d$-hyperboloid
\eqn\ekd{H^d={SO(d,1) \over SO(d)}}
(which has a Euclidean metric and was referred to in \witads\ as
Euclidean anti-de Sitter space) and the space
\eqn\eke{AAdS_d={SO(d-1,2) \over SO(d-2,2)}{},}
with two-timing signature $(d-2,2)$.

Here we will study the de Sitter and brane solutions of the M, string and
supergravity theories in
various signatures,
 and
investigate their properties. These give clues to the interpretation of these
theories, and also lead to generalisations of the conjectured duality between
3+1 dimensional $N=4$ super Yang-Mills theories and IIB string theory in
5-dimensional
anti-de Sitter space \mal, and between
  $N=4$ super Yang-Mills theories in 4 Euclidean dimensions and IIB*  string
theory in
5-dimensional
 de Sitter space \chrisone, as will be discussed elsewhere \nextpaper.

The plan of this paper is as follows: we outline in section 2
the type of brane and brane-related solutions in theories of
various spacetime signatures that we will investigate
in the rest of the paper.
In section 3 we write down the D-brane solutions of the type II
theories and the corresponding E-brane solutions of the II*
theories. We also display some wavelike solutions of the II* theories
that have no D-brane analogues. In section 4   we briefly review
the supersymmetric brane and Kaluza-Klein solutions of 10+1 dimensional
M-theory.
In section 5, we consider solutions of the eleven-dimensional M* theory
with signature 9+2 and discuss their geometry, while in
section 6, we discuss solutions of M$'$-theory with signature 6+5. In section
7, we summarize the
solutions of the various 11-dimensional M-theories and 10-dimensional type II
theories.
We discuss in section 8 the general
rules governing which world-volume signatures occur in which theories.
In section 9, we discuss the reductions of our eleven-dimensional
solutions to obtain solutions in ten dimensions, using this as
a check on the results of section 7. We also
show how D-branes and E-branes are related via
timelike reduction and oxidation with the specific example of the
D2-brane of IIA and the E1-brane of  IIB* following the methods
of \rust.

\newsec{Brane Solutions}

For theories in spacetimes of signature $(S,T)$, we
 will seek generalised brane solutions with metric   of the
form
\eqn\eki{ds^2=H^\aa \eta _{ab} dX^a dX^b + H^\bb \ti \eta _{ij} dY^i dY^j,}
where $\eta _{ab}$ is a flat metric of signature $(s,t)$ and
$\ti \eta _{ij}$ is a flat metric of signature $(S-s,T-t)$.
We will require $H$ to be a function of the transverse coordinates
$Y^i$ and find that the field equations imply that $H(Y)$   has to be a
harmonic function,
satisfying
\eqn\ekii{\ti \eta ^{ij} \pa _i \pa _j H=0,}
and determine the constants $\aa,\bb$.
The longitudinal space has signature $(s,t)$, and we refer to it as an
$(s,t)$-brane, so that a conventional
$p$-brane of a
Lorentzian theory with $(S,T)=(D-1,1)$ is a $(p,1)$-brane.
These solutions also have a non-vanishing $n$-form gauge field with $n=s+t$.

Different types of solutions arise for different choices of harmonic function.
A simple choice is the wave-type solution
\eqn\ekiii{H=A \sin (K_iY^i+c), \qq \ti \eta ^{ij}K_iK_j=0}
with $K$ a constant null vector.
For the  Euclidean transverse space (as in $p$-branes) there are no non-trivial
  null vectors $K$, but there are
non-trivial solutions
   if the transverse space has both spacelike and timelike dimensions.

A simple generalisation of the usual $p$-brane ansatz is to take
\eqn\ekiv{H=c +{Q\over  \vert Y \vert^q}, \qq q= S+T-s-t-2,}
for some constants $c,Q$; $c$ can be taken to be $0$ or $1$, while $Q$
represents the \lq charge' of the brane.
Here  $ \vert Y \vert ^q=\vert \ti \eta _{ij} Y^i Y^j\vert ^{q/2}$.
The solution then clearly has a potential singularity at $Y^2 =0$, 
where $Y^2 =\ti \eta _{ij} Y^i Y^j$;
this is the horizon at the point
$Y=0$ for the usual $p$-branes, but is more complicated in other signatures.
The null-cone $Y^2=0$ splits the transverse space
into two regions,
$Y^2>0$ and
$Y^2<0$, and in general a solution valid in one region cannot be continued
through the null cone to a solution
in the other region; in some examples, the surface $Y^2=0$ is   singular, in
others it is a boundary.
However, in the cases in which it is a non-singular horizon, one can continue
through it, as in {\inter}.
Near $Y^2=0$, the constant term in \ekiv\ is negligible, so that
one can take
\eqn\ekv{H \sim {Q\over  \vert Y \vert^q}.}
This limiting form of the geometry is sometimes a non-singular product of de
Sitter-type spaces.
For example, for the usual D3-brane, M2-brane and M5-brane it is a product of
anti-de Sitter space and a
sphere, $AdS_{p+2}\times S^{D-p-2}$ \refs{\inter,\mal}, while for
the E4-brane of the type IIB* string theory, it is
the product $dS_5\times H^5$
of 5-dimensional de Sitter space and a 5-hyperboloid \chrisone.
For the case of the E4-brane, the two regions $Y^2>0$ and $Y^2<0$ are two
distinct, complete, non-singular
solutions and for both cases  $Y^2=0$ is a boundary at infinite distance.
There are thus two solutions, and neither can be analytically continued through
$Y^2=0$.
The interpretation of these solutions was discussed in \chrisone.

In all the cases with Lorentzian signature transverse spaces, we will exhibit
two solutions, one
for the region $Y^2>0$ and one for the region $Y^2<0$. We will not address here
the nature of the
surface $Y^2=0$ (which could be a singularity, a horizon, a boundary, etc.) or
discuss in which cases the brane solutions can
be continued through $Y^2=0$. We hope to return to these questions, and others
concerning the interpretation of these
solutions, in a future publication.

In addition to the single source solution \ekiv,
there are multi-source solutions
\eqn\ekmult{H=c +\sum _m{Q_m\over  \vert Y-Y_m \vert^q}}
as well as smeared solutions in which the harmonic function $H$ is independent
of one or more of the $Y^i$.
However, in each case we will usually only display the single-source solution
\ekiv\ explicitly.

\newsec{D-branes and E-branes}

The bosonic action of the IIA supergravity is
\eqn\twoa{S_{IIA}=\int d^{10} x \sqrt{-g}\left[
 e^{-2 \Phi}\left(R+ 4(\partial   \Phi )^2  -{H^2\over 12}\right)
-{G_2^2\over 4} - {G_4^2\over 48} \right] +
{4\over \sqrt 3}\int G_4 \wedge G_4 \wedge B_2 + \dots }
while that of IIB supergravity is
\eqn\twob{S_{IIB}=\int d^{10} x \sqrt{-g}\left[
 e^{-2 \Phi} \left( R+ 4(\partial   \Phi )^2
- {H^2\over 12} \right)-{G_1^2\over 2} -  {G_3^2\over 12}
- {G_5^2\over 240}\right] + \dots}
Here  $\Phi $ is the dilaton, $H=dB_2$ is the field strength of the
NS-NS 2-form gauge field $B_2$ and
$G_{n+1}=dC_n+\dots$ is the field strength for the RR $n$-form gauge field
$C_n$. The field
equations derived from the IIB action \twob\ are
supplemented with the self-duality constraint $G_5=*G_5$.
Our conventions are that in signature $S+T$, the metric has $S$ positive
spatial eigenvalues
and $T$ negative timelike ones, so that
a Lorentzian metric has signature $(++\dots +-)$.

The  type II supergravity solution for a D$p$-brane ($p$ is even
for IIA and odd for IIB) is given by
\refs{\hor,\prep,\huten}
\eqn\dbr{\eqalign{
 ds^2&=H^{-1/2}(-dt^2+dx_1^2+\dots+dx_p^2)+H^{1/2}(dy_{p+1}^2+\dots+dy_9^2)\cr
e^{-2\Phi}&=H^{(p-3)/2}, \qquad\qquad C_{012\dots p}=-H^{-1}+k,\cr}}
where $H$ is a harmonic function of the transverse coordinates
$y_{n+1},\dots,y_9$, $k$ is a constant and here and throughout in the
paper we denote longitudinal spatial coordinates by $x_a$ and transverse
spatial coordinates by $y_i$.
The simplest choice for $H$ is
\eqn\hre{H=c+ {q\over y^{7-p}},}
where $c$ is a constant (which can be taken to be $0$ or $1$), $q$ is the
D-brane charge and $y$ is
the radial coordinate defined by
\eqn\abc{y^2=\sum_{i=p+1}^9 y_i^2.}
When $c\neq 0$, it is conventional to set $k=c^{-1}$, so that as
$y \to \infty$, $C_{012\dots p} \to 0$. However, for convenience we will
henceforth set $k=0$ and usually take $c=1$.

The type IIA* and type IIB* actions are given by reversing
the signs of the RR kinetic terms in \twoa,\twob\ to give \chrisone
\eqn\twoas{S_{IIA*}=\int d^{10} x \sqrt{-g}\left[
 e^{-2 \Phi} \left(  R+ 4(\partial   \Phi )^2
-{H^2\over 12} \right)+{G_2^2\over 4} +{G_4^2\over 48}\right] + \dots}
and
\eqn\twobs{S_{IIB*}=\int d^{10} x \sqrt{-g}\left[
 e^{-2 \Phi} \left( R+ 4(\partial   \Phi )^2
- {H^2\over 12}  \right)+{G_1^2\over 2} +  {G_3^2\over 12}
+ {G_5^2\over 240} \right] + \dots,}
where the field equations from \twobs\ are supplemented by the constraint
$G_5=*G_5$.

The E$p$-brane solutions to \twoas\ and \twobs\ are given by
\eqn\ebr{\eqalign{
 ds^2&=H^{-1/2}(dx_1^2+\dots+dx_p^2)+H^{1/2}
(-dt^2+dy_{p+1}^2+\dots+dy_9^2),\cr
e^{-2\phi}&=H^{(p-4)/2}, \qquad \qquad C_{12\dots p}=-H^{-1}.\cr}}
In this case, $H$ is a harmonic function of
$t,y_{p+1},\dots,y_9$
(i.e. it is a solution of the
wave equation $\nabla ^2
H=0$), and $H$ can depend on time as
well as the spatial transverse coordinates.

There are a number of different possibilities for $H$. First, we can take the
time-independent $H$ given by \hre.
These are the solutions that arise from the D-brane supergravity solutions on
performing a timelike
T-duality, using a generalisation \christwo\ of Buscher's rules \buscher.
Secondly, we can consider the  solution  \chrisone\
\eqn\ehar{ H=c+ {q\over \tt ^{8-p}}}
where $\tt, \ss $ are the proper time and distance defined by
\eqn\abcd{\tt^2 =  -\ss ^2=  t^2- y^2.}
This corresponds to a source located at a point in the transverse space-time.
For odd $p$, taking
\eqn\ehars{H=c+ {q\over \ss ^{8-p}}}
gives a different solution.

The solution \ebr\ is an extended object associated with a spacelike
$p$-surface with
coordinates $x^1,\dots ,x^p$
located at $t=y^{p+1}=\dots =y^9=0$.
This is to be compared with  a D-brane, which is associated
 with a timelike $p+1$-surface with coordinates $t,x^1,\dots ,x^p$
located at $ y^{p+1}=\dots =y^9=0$.
A D-brane arises in perturbative type II string theory from imposing
Dirichlet boundary conditions
in the directions $ y^{p+1},\dots ,y^9 $ and Neumann conditions in the
remaining directions, and the
D-brane solution \dbr\ describes the supergravity fields resulting from
such a D-brane source.
Similarly, the E-brane in perturbative type II* string theory
arises from imposing Dirichlet
boundary conditions in the directions $t, y^{p+1},\dots ,y^9 $, including
time, and Neumann conditions
in the remaining directions, and the E-brane solution \ebr\
describes the supergravity fields resulting
from such an E-brane source. In the perturbative string theory, the
timelike T-duality taking the
type II theory to the type II* theory changes the boundary condition
in the time direction from
Dirichlet to Neumann, and so takes a D$p$-brane to an E$p$-brane.
The  solutions have a potential  singularity on the light-cone $t^2=y^2$,
but in some cases this is non-singular \chrisone.
The E-branes preserve 16 of the 32 supersymmetries of the type II*
theories.
Smearing these solutions in the time direction gives the time-independent
solutions given by \ebr\ with $H$ given by \hre, and other solutions can be
obtained by smearing in spacelike directions.

Another choice of $H$ satisfying the wave equation is a wave solution
\eqn\wave{H (y^i,t)=A \sin (k_iy^i+\omega t), \qquad k^2=\omega ^2}
or a linear supposition of such solutions.
These are completely non-singular and have no non-trivial D-brane analogue.

\newsec{Solutions of M-Theory}

M-theory is the strong coupling limit of the type IIA string \wit\ and is a
theory in 10+1 dimensions
whose low-energy effective field theory is 11-dimensional
supergravity \cjs\ with bosonic action
\eqn\mmm{S_M=\int d^{11} x \sqrt{-g} \left( R - {G_4^2\over 48}
\right) -{1\over 12} \int C\wedge G \wedge G.}
In this section, we review the solutions of M-theory
whose analogues for M* and M$'$ theories we will find in later sections.

The maximally supersymmetric solutions are 11-dimensional Minkowski space
and its toroidal compactifications,
the Freund-Rubin solution $AdS_4\times S^7$ \fr, and the compactification
$AdS_7\times S^4$ \pvt.
It also has solutions which preserve half
of the 32 supersymmetries: the M2-brane  \dufstel, the M5-brane \guven, the
pp-wave solution \bko\ and the
Kaluza-Klein (KK) monopole \gps\ solution
$\R^{6,1}\times TN$ {\PKT}, where $\R^{s,t}$ is flat space with  signature
$(s,t)$,
and TN is the self-dual Euclidean Taub-NUT solution with metric
\eqn\TN{ds^2=V^{-1} \left(dz+A_i dy^i\right)^2+V\delta_{ij}dy^i dy^j,}
where $V$ is a harmonic function depending on $y_i$ and
\eqn\tnt{F_{ij} \equiv \partial_i A_j - \partial_j A_i=\epsilon_{ijk}
\partial_k V.}

The M2-brane solution of the 10+1 supergravity action \mmm\ is given by
\dufstel\
\eqn\mtwo{\eqalign{ds^2=&H^{-2/3} (-dt^2+dx_1^2+dx_2^2)
+ H^{1/3}(dy_3^2 + \dots + dy_9^2+dy_{10}^2),\cr
C_{012}&=H^{-1},\cr}}
where $H(y_3,\dots ,y_{10})$ is a harmonic function
in the transverse space. For a single membrane at $y=0$, we take
\eqn\fdkfo{H=1+{Q\over y^6},}
 where here and throughout $y^2=\sum_i y_i^2$, where $i$ runs over
the spatial indices in the transverse space.
The world-volume has signature (2,1).
The M2-brane solution has bosonic symmetry $ISO(2,1) \times SO(8)$.
It is nonsingular at $y=0$ with near-horizon geometry
$AdS_4 \times S^7$ \inter.

The   M5-brane   \guven\ is given by
\eqn\mfive{\eqalign{ds^2&=H^{-1/3} (-dt^2+dx_1^2 + \dots  +dx_5^2)
+H^{2/3} (dy_6^2 + dy_7^2+dy_8^2+dy_9^2+dy_{10}^2),\cr
&C_{t12345}=H^{-1},\cr}}
where
\ek{H=1+{Q \over y^3}}
 and $y^2=y_6^2+y_7^2 + \dots + y_{10}^2$.
This solution has bosonic symmetry $ISO(5,1) \times SO(5)$
and interpolates between $AdS_7 \times S^4$ and flat space \inter.

\newsec{Solutions of M*}

We will now seek the analogues of the M2-brane, M5-brane and KK monopole that
occur in the M* theory.
The M* theory
\christwo\ is  the strong coupling limit of  the IIA* theory and is a theory
in 9+2 dimensions
whose field theory limit is a supergravity theory with bosonic action \chrisone
\eqn\mstar{S_M=\int d^{11} x \sqrt{g} \left( R + {G_4^2\over 48} \right)
-{1\over 12} \int C \wedge G \wedge G.}
Note that the sign of the kinetic term of  $G_4$ is opposite to that of the
action \mmm;
as we shall see in section 8, the sign of the kinetic term is intimately
related with the
world-volume signatures that can occur.
For example, if the sign of the kinetic term of $G_4$ were reversed in \mstar\
to give a Lagrangian
$R-  {G_4^2/ 48}+...$ in 9+2 dimensions, there would be a membrane solution
with 2+1
dimensional world-volume,
while the
action \mstar\ with the opposite sign for the $G_4$ kinetic term has brane
solutions
with world-volume signatures (3,0) and (1,2), as we shall see below.
The sign of the $G_4$ kinetic term in actions \mmm,\mstar\ is determined by
supersymmetry.

The equations of motion of this action are straightforward to
solve, using Appendix B for the Ricci tensor and scalar.
This theory has a number of Freund-Rubin-type solutions involving
the de-Sitter-type spaces
defined in \eka\ -- \eke, including $d$-dimensional
de Sitter space $dS_d$,  $d$-dimensional
anti-de de Sitter space $AdS_d$, the  $d$-dimensional
hyperbolic space $H_d$, and the two-time de Sitter space
$AAdS_{d}$ which is a generalised  de Sitter space given by
(a connected component of) the
coset
$SO(d-1,2)/SO(d-2,2)$, with signature $(d-2,2)$ and isometry $SO(d-1,2)$.
The solutions  are
\eqn\geoms{\eqalign{
& dS_4  \times AdS_7
\cr &
AAdS_4   \times S^7
\cr &
AdS_7
\times dS_4
\cr &
AAdS_7  \times H^4
\cr}}
For example, the
$AAdS_4\times S^7$ solution is
\eqn\mscomp{\eqalign{ds^2&={z^2 \over a^2} (-dt^2-dt'^2+dx^2) + { a^2\over
z^2}dz^2+
4a^2 d\Omega_7^2,\cr
&C_{tt'xz}={3\over a} \epsilon_{tt'xz}.\cr}}
$AAdS_4$ has signature $(2,2)$,  isometry $SO(3,2)$ and constant curvature
$\sim 1/a^2$.

There are two solutions of M*-theory analogous to the M2-brane.
The first of these is the (1,2)-brane given by
\eqn\msonetwo{\eqalign{ds^2=&H^{-2/3} (-dt^2-dt'^2+dx^2)
+ H^{1/3}(dy_2^2 + \dots + dy_9^2),\cr
C_{tt'x}&=H^{-1},\cr}}
where $H(y_3,\dots ,y_{10})$ is again a  harmonic function in $\R^8$,
which we can take to be \fdkfo. The world-volume has signature
(1,2), with two times. This solution has bosonic symmetry
$ISO(1,2) \times SO(8)$.
Near $y=0$, the metric takes the form
\eqn\msonetwomet{ds^2 = {U^2\over R^2} (-dt^2-dt'^2+dx^2) + {R^2dU^2\over U^2}
+Q^{1/3}d\Omega_7^2,}
which is the metric on $AAdS_4\times S^7$; here $U=Q^{-1/6}y^2/2$ and
$R=Q^{1/6}/2=R_{S^7}/2$.
The (1,2)-brane interpolates between the flat space $\R^{9,2}$ and
$AAdS_4\times S^7$.

The second membrane-type solution is the (3,0)-brane given by
\eqn\msthreezero{\eqalign{ds^2=&H_2^{-2/3} (dx_1^2+dx_2^2+dx_3^2)
+ H_2^{1/3}(-dt^2-dt'^2+dy_4^2 + \dots + dy_9^2),\cr
C_{123}&=H_2^{-1},\cr}}
where $H$ is a harmonic function  on the transverse space.
The world-volume is Euclidean, with
signature (3,0).

The E4-brane solution considered in \chrisone\
has a transverse space which is
6-dimensional Minkowski space, which is divided into two regions by the
light-cone.
Choosing the harmonic function \ehar\ gave a solution which split into two
regions, the inside and
outside of the light-cone, and each region is in fact a non-singular complete
solution, with the
light-cone of the parameter space becoming a boundary at infinite geodesic
distance.
A similar situation occurs here for the (3,0)-brane.

For the (3,0) brane, the null-cone $y^2=t^2+t'^2$ divides the transverse
parameter space into two
regions,
and, as discussed in section 2, there are two distinct brane solutions,
in which the transverse coordinate space is
restricted to the region inside or outside the null cone.
In the region
$y^2> t^2+t'^2$,
a natural choice for the time-dependent harmonic
function is
\eqn\tryrea{H_2=1+{Q\over (y^2-t^2-t'^2)^3},}
which gives a real solution  \msthreezero\ for $y^2> t^2+t'^2$.
For $y^2< t^2+t'^2$, we take instead
\eqn\tryreb{H_2=1+{Q\over (t^2+t'^2-y^2)^3}.}
In either case, the solution has bosonic symmetry
$ISO(3) \times SO(6,2)$.
The geometry of \msthreezero\ near $y^2=t^2+t'^2$
differs in the two cases, and the interpretation is similar to that of the
E4-brane solution of the  IIB* string  \chrisone.

For $y^2 > t^2 + t'^2$, let $\sigma^2=y^2-t^2-t'^2$.
Then near $\sigma =0$, $H_2^{1/3}\sim Q^{1/3}/\sigma^2$.
Setting
\ek{y=\sigma \cosh \alpha ,
\qq t=\sigma \sinh\alpha\cos\beta, \qq t'=\sigma \sinh\alpha\sin\beta}
it is straightforward to show that, near $\sigma=0$ the metric takes the form
\eqn\msthreezeroa{ds^2 ={V^2\over R^2} (dx_1^2+dx_2^2+dx_3^2) +
{R^2 dV^2\over V^2}+Q^{1/3} \left(
-d\alpha^2 - \sinh^2\alpha d\beta^2  + \cosh^2\alpha d\Omega_5^2 \right),}
where $V=Q^{-1/6} \sigma^2/2$ and again $R=Q^{1/6}/2$. This
is the metric for  $H^4 \times AAdS_7$.
The region $y^2 > t^2 + t'^2$ of the solution \msthreezero\  then
interpolates between the flat space $\R^{9,2}$
and  $H^4 \times AAdS_7$.

For $t^2+t'^2 > y^2$, let $\tau^2=t^2+t'^2-y^2$.
Then near $\tau=0$, $H_2^{1/3}\to Q^{1/3}/\tau^2$.
Setting
\ek{ y=\tau \sinh \alpha, \qq
 t=\tau \cosh\alpha\cos\theta, \qq t'=\tau \cosh\alpha\sin\theta ,}
 the metric near $\tau=0$ takes the form
\eqn\msthreezerob{ds^2 ={W^2\over R^2} (dx_1^2+dx_2^2+dx_3^2) -
{R^2dW^2\over W^2} + Q^{1/3}\left(
d\alpha^2 -  \cosh^2\alpha d\theta^2 + \sinh^2\alpha d\Omega_5^2\right),}
where $W=Q^{-1/6} \tau^2/2$ and again $R=Q^{1/6}/2$.
This is the metric of  $dS_4 \times AdS_7$.
The region $t^2+t'^2 > y^2$ of the solution \msthreezero\
then interpolates between the flat space $\R^{9,2}$
and $dS_4 \times AdS_7$.

The M* theory has a  (5,1)-brane  solution (analogous to the M5-brane of
M-theory)
 which is given by
\eqn\mfivestar{\eqalign{ds^2&=H^{-1/3} (-dt^2+dx_1^2 + \dots  +dx_5^2)
+H^{2/3} (-dt'^2 +dy_6^2 + dy_7^2+dy_8^2+dy_9^2),\cr
&C_{t12345}=H^{-1},\cr}}
Here $H$ is a harmonic function and
again there are two solutions.
In the first solution
\eqn\ewtrw{H=1+{Q\over\tau^3},}
where the transverse coordinates are restricted to the region
 $\tau^2=y^2-t'^2 > 0$, where   $y^2=y_6^2+y_7^2 + \dots + y_9^2$.
In the other solution
\eqn\ufgjh{H=1+{Q\over\sigma^3},}
where the transverse coordinates are restricted to the region
 $\sigma^2=t'^2-y^2>0$.
Both solutions have bosonic symmetry
$ISO(5,1) \times SO(4,1)$ and world-sheet signature (5,1).

Again, the geometry differs for the two cases. In the first
case, near $\tau=0$, it takes the form of $AAdS_7 \times H^4$,
while for the second, near $\sigma=0$ it takes the form
$AdS_7 \times dS_4$. In both cases, the brane solution interpolates
between these geometries and flat space.
Note that these near horizon geometries are the same
as those for the Euclidean membrane with world-volume signature $(3,0)$.

There are also generalisations  of the
KK-monopole solution of M-theory.
One is $\R^{5,2} \times TN$, where $TN$ is the
Euclidean self-dual four-dimensional Taub-NUT space \TN, while the others
involve  $
 TN^{2,2}$, where $TN^{2,2}$  is
the generalisation of Taub-NUT space with signature 2+2
given by \bgppr\
\eqn\TNtwo{ds^2=\left(-V^{-1} \left(dz+A_i dY^i\right)^2
+V\eta_{ij}dY^i dY^j\right),}
where here $\eta_{ij}$ is a flat
metric with signature $( +,+,-)$, $i,j=1,2,3$.
$V$ is again a  harmonic function depending on $Y_i$
and $curl A = grad V$, where, following \bgppr,
$curl$ and $grad$ are defined with respect to the Lorentizian metric
$\eta_{ij}$.
Writing $Y^i=(y_1,y_2,t)$, $Y^2 =\eta_{ij} Y^i  Y^j= y^2-t^2$, the harmonic
function can be chosen as
\ek{
V= c + {Q\over Y}= c + {Q\over ( y^2-t^2) ^{1/2}}}
in the region $Y^2>0$, or as
\ek{
V=   c + {Q\over ( t^2-y^2) ^{1/2}}}
in the region $Y^2<0$.
For the $TN^{2,2}$ solution, we take $c=1$.
The timelike coordinate $z$ is periodic, with period $4\pi Q$.
Reversing the signature of the metric gives another space of signature (2,2)
with metric
\eqn\TNtwoa{ds^2=-\left(-V^{-1} \left(dz+A_i dY^i\right)^2
+V\eta_{ij}dY^i dY^j\right),}
in which space and time have been interchanged, so that in particular  the $z$
coordinate on the $S^1$ fibre is now
spacelike; we shall denote this space as $-TN^{2,2}$. There are then two
distinct solutions,
$\R^7\times TN^{2,2}$  and
$\R^7\times -TN^{2,2}$.

Both $TN$ and $TN^{2,2}$ are self-dual Kahler Ricci-flat spaces with two
covariantly constant
spinors;
$TN$ has holonomy $SU(2)$ while $TN^{2,2}$  has holonomy $SU(1,1)$.
Then the solutions  $\R^{5,2} \times TN$, $\R^7\times TN^{2,2}$  and
$\R^7\times -TN^{2,2}$ preserve half the supersymmetry.
There is also a Lorentzian signature Taub-NUT solution $TN^{3,1}$ \bgppr,
but this has no covariantly
constant spinors and so the  solutions   $\R^7\times TN^{3,1}$ of M-theory and
$\R^{6,1}\times TN^{3,1}$ of M* theory do not preserve any supersymmetries, and
will not be
considered further here; such solutions have been considered in \chamblin.
There are also supersymmetric pp-wave solutions similar to those of \bko.

\newsec{Solutions of M$'$-Theory}

There is a IIA string theory in a spacetime with signature 5+5 whose strong
coupling limit is the M$'$ theory with signature 6+5
\refs{\christwo}.
The field theory limit of M$'$ theory
is a supergravity theory in 6+5 dimensions with bosonic action
 \eqn\mprime{S_{M'}=\int d^{11} x \sqrt{-g} \left( R - {G_4^2\over 48}
\right) -{1\over 12} \int C\wedge G \wedge G.}

There are the following Freund-Rubin-type
 solutions:
\eqn\ekfr{\eqalign{
AAdS_4& \times {SO(4,4)\over SO(4,3)} ={SO(3,2)\over SO(2,2)}  \times
{SO(4,4)\over SO(4,3)}
\cr
AdS_4 & \times {SO(4,4)\over SO(3,4)} ={SO(3,2)\over SO(3,1)}  \times
{SO(4,4)\over SO(3,4)}
\cr
-dS_4 &
\times   AAdS_7 ={SO(1,4)\over SO(1,3)}  \times {SO(6,2)\over SO(5,2)}
\cr
-H^4  & \times AdS_7 ={SO(1,4)\over SO(4)}   \times {SO(6,2)\over SO(6,1)}
\cr
S^4 & \times   -AAdS_7 ={SO(5)\over SO(4)}   \times {SO(2,6)\over SO(2,5)}
\cr}}
Here if $X$ denotes a space with metric $g$ and signature $(p,q)$, then
$-X$ denotes
the same space $X$, but with   metric $-g$ and signature $(q,p)$.
Then $-dS_4 = SO(1,4)/SO(1,3)$ has signature (1,3) and
  $-H^4=SO(1,4)/SO(4)$ is a timelike hyperboloid with  negative definite metric
of signature
(0,4).

There are the following half-supersymmetric KK-monopole type solutions
\eqn\ekkk{\eqalign{
\R^{2,5} &\times  TN   \cr
\R^{4,3} &\times TN ^{2,2} \cr
\R^{6,1} &\times -TN   \cr}}
and  pp-wave solutions similar to those of {\bko}.

There are two solutions analogous to the M2-brane. The first is the
(2,1)-brane
\eqn\mptwoone{\eqalign{ds^2=&H^{-2/3} (-dt^2+dx_1^2+dx_2^2)
+ H^{1/3}(-d\ti t_3^2-d\ti t_4^2-d\ti t_5^2-d\ti t_6^2
+dy_7^2 + \dots + dy_{10}^2),\cr
C_{t12}&=H^{-1}.\cr}}
The null cone splits the transverse space into two regions, and there   is a
solution for each
region.
For the region
\eqn\ektoa{\ti t^2=\ti t_3^2+\dots +\ti t_6^2 > y^2=y_7^2+\dots +y_{10}^2}
we take the harmonic function
\eqn\ektoah{H=1+{Q\over(\ti t^2-y^2)^3},}
while
for
\eqn\ektob{ y^2 > \ti t^2}
we take
\eqn\ektobh{H=1+{Q\over(y^2-\ti t^2)^3}.}
Both solutions have bosonic
symmetry $ISO(2,1) \times SO(4,4)$ and the first solution \ektoah\ interpolates
between
$AAdS_4 \times SO(4,4)/SO(4,3)$ and  $\R^{6,5}$ while  the second \ektobh\
interpolates  between
  $AdS_4 \times SO(4,4)/SO(3,4)$ and $\R^{6,5}$.

The second solution anaogous to the M2-brane is the (0,3)-brane given by
\eqn\mpzerothree{\eqalign{ds^2=&H_2^{-2/3} (-dt_1^2-dt_2^2-dt_3^2)
+ H_2^{1/3}(-d\ti t_1^2-d\ti t_2^2+dy_4^2 + \dots + dy_9^2),\cr
C_{t_1t_2t_3}&=H_2^{-1}.\cr}}
Again there are two regions and a natural choice  for the harmonic function in
each, giving two
distinct solutions:
\eqn\ekzta{H_2=1+{Q\over
(y^2-w^2)^3}}
  for $y^2=y_4^2+\dots +y_9^2 > \ti t^2=\ti t_1^2+\ti t_2^2$
or
\eqn\ekztb{
H_2=1+{Q\over (\ti t^2-y^2)^4}}
 for $\ti t^2 > y^2$.
It has bosonic symmetry $ISO(3) \times SO(6,2)$
which is identical to that of \msthreezero, although
spatial and temporal dimensions are interchanged in the worldvolume.
\mpzerothree\ interpolates between $-dS_4 \times AAdS_7$
and $\R^{6,5}$ in the first case and between $-H_4 \times AdS_7$
and $\R^{6,5}$ in the second.

Three solutions of \mprime\ analogous to the M5-brane can be
obtained. The first is the (5,1)-brane given by
\eqn\mpfiveone{\eqalign{ds^2&=H^{-1/3} (-dt^2+dx_1^2 + \dots  +dx_5^2)
+H^{2/3} (-d\ti t_6^2-d\ti t_7^2-d\ti t_8^2-d\ti t_9^2 + dy_6^2),\cr
&C_{t12345}=H^{-1},\cr}}
where
\eqn\ekfoa{
H=1+{Q\over
\tau^3}} for $\tau^2=\ti t^2-y_6^2 > 0$ and
\eqn\ekfob{H=1+{Q\over \sigma^3}}
for $\sigma^2=y_6^2-\ti t^2 > 0$, where $\ti t^2=\ti t_6^2+\dots +\ti t_9^2$.
This solution
has bosonic symmetry $ISO(5,1) \times SO(4,1)$.
\mpfiveone\ interpolates between $-dS_4 \times AAdS_7$
and $\R^{6,5}$ in the first case and between $-H^4 \times AdS_7$
and $\R^{6,5}$ in the second.

The (3,3)-brane solution is given by
\eqn\mpthreethree{\eqalign{ds^2&=H^{-1/3} (-dt_1^2-dt_2^2-dt_3^2+
dx_1^2 + dx_2^2+dx_3^2)
+H^{2/3} (-d\ti t_1^2-d\ti t_2^2+dy_1^2+dy_2^2+dy_3^2)\cr
&C_{t_1t_2t_3x_1x_2x_3}=H^{-1},\cr}}
where
\eqn\ektta{H=1+{Q\over \tau^3}}
 for $\tau^2=\ti t^2-y^2 > 0$ and
\eqn\ekttb{
H=1+{Q\over\sigma^3}}
for $\sigma^2=y^2-\ti t^2 > 0$, where $\ti t^2=\ti t_1^2+\ti t_2^2$ and
$y^2=y_1^2+y_2^2+y_3^2$. This solution has bosonic symmetry
$ISO(3,3)\times SO(3,2)$.
\mpthreethree\ interpolates between $SO(4,4)/SO(3,4) \times AdS_4$
and $\R^{6,5}$ in the first case and between $SO(4,4)/SO(4,3)\times AAdS_4$
and $\R^{6,5}$ in the second.

The third solution of M$'$ analogous to M5 is the (1,5)-brane given by
\eqn\mponefive{\eqalign{ds^2&=H^{-1/3} (-dt_1^2 -dt_2^2- \dots
-dt_5^2 + dx_1^2)
+H^{2/3} (dy_2^2+dy_3^2+dy_4^2+dy_5^2+ dy_6^2),\cr
&C_{t_1t_2t_3t_4t_5x_1}=H^{-1},\cr}}
where
\eqn\ekof{H=1+{Q\over y^3},}
where $y^2=y_2^2+\dots +y_6^2$. This solution
has bosonic symmetry $ISO(5,1) \times SO(5)$. \mponefive\ interpolates between
$-AAdS_7\times S^4$
and $\R^{6,5}$.

\newsec{10-dimensional and 11-dimensional solutions}

In the previous sections, we have considered the branes that occur in
the M-theories in signature 10+1, 9+2 and 6+5. They are summarised in the
following table:

\vskip 0.5cm
{\vbox{
\begintable
  | $C_3$, $s+t=3$ | $\tilde C_6$, $s+t=6$ \elt
 $M_{10,1}$ | (2,1) | (5,1) \elt
 $M_{9,2}$ | (3,0), (1,2) | (5,1) \elt
 $M_{6,5}$ | (2,1), (0,3) | (5,1), (3,3), (1,5)
\endtable

{\bf Table 1} The M-branes with world-sheet signature $(s,t)$ coupling to 
$C_3$ or its dual $\ti C_6$ in the
various M-theories with signature
$(S,T)$.
}}
\vskip .5cm

There are also equivalent mirror theories with signatures (1,10), (2,9) and
(5,6) and these have
brane solutions with the reversed world-volume signatures. For example, the
theory in signature (2,9) has branes of signature (2,1), (0,3) and (1,5).

A similar analysis can be done for the versions of the IIA and IIB theories in
various signatures;
there are branes of signature $(s,t)$ coupling to $n$-form gauge fields
with $s+t=n$, where the $n$-form gauge fields consist of $B_2$ and its dual
$\ti B_6$, and
the RR gauge fields $C_n$ ($n \le 4$) and their duals $\ti C_{8-n}$.
The results are summarised in the following tables.

\vskip 1cm
{\vbox{
\begintable
  | $C_1$ | $B_2$ | $C_3$ | $\tilde C_5$ | $\tilde B_6$ | $\tilde C_7$ \elt
 $IIA_{10,0}$ | (1,0) | (2,0) | -- | (5,0) | -- | -- \elt
 $IIA_{9,1}$ | (0,1) | (1,1) | (2,1) | (4,1) | (5,1) | (6,1) \elt
 $IIA_{9, 1}^*$ | (1,0) | (1,1) | (3,0) | (5,0) | (5,1) | (7,0) \elt
 $IIA_{8,2}$ | (0,1) | (2,0),(0,2) | (3,0),(1,2) | (4,1) | (5,1) | (7,0),(5,2)
\elt
 $IIA_{6,4}$ | (1,0) | (2,0),(0,2) | (2,1),(0,3) | $\hbox{(5,0),(3,2),} \atop
\hbox{(1,4) }$| (5,1),(3,3) | (6,1),(4,3) \elt
 $IIA_{5,5}$ | (0,1) | (1,1) | (2,1),(0,3) | $\hbox{(4,1),(2,3),} \atop
\hbox{(0,5)}$ | $\hbox{(5,1),(3,3),} \atop
\hbox{(1,5) }$| (4,3),(2,5)
\endtable

{\bf Table 2} The branes with world-sheet signature $(s,t)$ of the various
$IIA_{S,T}$ theories with
signature
$(S,T)$.
}}
 \vskip 1cm
{\vbox{
\begintable
  | $C_0$ | $B_2$ | $C_2$ | $ C_4$ | $\tilde B_6$ | $\tilde C_6$ | $\tilde C_8$
\elt
 $IIB_{9,1}$ | -- |  (1,1) |  (1,1) | (3,1) | (5,1) | (5,1) | (7,1) \elt
 $IIB^*_{9,1}$ | (0,0) | (1,1) | (2,0) | (4,0) | (5,1) | (6,0) | (8,0) \elt
 $IIB_{9, 1}'$ | (0,0) |  (2,0) |  (1,1) | (4,0) | (6,0) | (5,1) | (8,0) \elt
 $IIB_{7,3}$ | -- |  $\hbox{(2,0),} \atop \hbox{(0,2) }$|  $\hbox{(2,0),} \atop
\hbox{(0,2)}$ |
(3,1),(1,3) | (6,0),(4,2) | (6,0),(4,2) |
$\hbox{(7,1),} \atop \hbox{(5,3) }$\elt
 $IIB_{5,5}$ | -- | (1,1) |  (1,1) |  (3,1),(1,3) | $\hbox{(5,1),(3,3),} \atop
\hbox{(1,5)}$ |
$\hbox{(5,1),(3,3),} \atop \hbox{(1,5)}$ |
$\hbox{(5,3),} \atop \hbox{(3,5) }$\elt
 $IIB^*_{5,5}$ | (0,0) | (1,1) |  $\hbox{(2,0),} \atop \hbox{(0,2)}$ |
$\hbox{(4,0),(2,2),} \atop
\hbox{(0,4)}$ |
$\hbox{(5,1),(3,3),} \atop \hbox{(1,5)}$ | (4,2),(2,4) | (4,4) \elt
 $IIB_{5,5}'$ | (0,0) | $\hbox{ (2,0),} \atop \hbox{(0,2) }$|  (1,1) |
$\hbox{(4,0),(2,2),} \atop
\hbox{(0,4)}$ | (4,2),(2,4) | $\hbox{(5,1),(3,3),} \atop \hbox{(1,5)}$ | (4,4)
\endtable

{\bf Table 3} The branes with world-sheet signature $(s,t)$ of the various
$IIB_{S,T}$ theories with
signature
$(S,T)$.
}}

The IIB theories also have D-instanton-type solutions, which will have real or
imaginary
$C_0$ charge, depending on how the theory is Wick-rotated; see {\chrisone}.

 \vskip .5cm

\newsec{Solutions vs. Signature}

In this section, we will systematically consider a generic theory with
spacetime signature $(S,T)$
and investigate  which branes can occur with which world-volume signatures.
Consider a D-dimensional action  of the form
\eqn\mgen{S_M=\int d^D x \sqrt{\vert g \vert} \left( R -\epsilon e^{a \phi}
F_{n+1}^2 -{1\over 2} (\nabla \phi)^2\right) +\dots,}
where $F_{n+1}$ is the field  strength for an $n$-form gauge field $A_n$,
with $n \le D-2$,
$\epsilon=\pm 1$, and we have included a possible dilaton coupling
$e^{a \phi}$ in the kinetic term
for $A$; in the 11-dimensional theories, this will be absent ($a=0$).
We have allowed an arbitrary sign $\epsilon=\pm 1$ for the gauge field kinetic
term, and will
examine how the choice of sign affects the
brane solutions, and in particular their world-volume signature.

The spacetime signature is $(S,T)$, $S+T=D$ and
we will
 seek brane-type solutions with metrics of the form
\eqn\ertsgf{ds^2=A dX^2 + BdY^2,}
where the   longitudinal space has coordinates $X^1,\dots X^d$ and flat metric
$dX^2$   with signature $(s,t)$, $s+t=d$, and the
transverse  space has coordinates $Y^1,\dots Y^{D-d}$ and flat metric
$dY^2$   with signature $(S-s,T-t)$;
$A,B$ are functions of
$Y$. Without loss of generality, we take $S\ge T$.
Note that dualising the $n$-form $A_n$ to a $\ti n=D-n-2$ form
$\tilde A_{\ti n}$
with field strength
$\ti F_{\ti n+1} = e^{-a\phi} * F_{n+1}$ (when possible)
gives a dual form of the action
\eqn\mgendu{S_M=\int d^D x \sqrt{\vert g \vert} \left( R -\ti
\epsilon e^{-a \phi} \ti
F_{\ti n+1}^2 -{1\over 2} (\nabla \phi)^2\right) +\dots,}
where
\eqn\abc{\ti \epsilon = (-1) ^{T-1}\epsilon.}

With Lorentzian signature $(S,T)=(D-1,1)$ and the conventional sign
$\epsilon =1$, the $n$-form
gauge field couples to   an electrically charged
$p$-brane with $p=n-1$ and world-volume signature $(p,1)$
and to a
 magnetically charged
$p'$-brane with $p'=D-n-3$ and world-volume signature $(p',1)$,
and these are the only such brane
solutions that arise.
If the sign of the $A_n$ kinetic term is reversed, so that $\epsilon =-1$,
there are now
Euclidean brane solutions, where the electrically charged
brane has world-volume signature $(n,0)$ (provided $n\leq S$)
and the magnetically charged
brane has world-sheet signature $(D-n-2,0)$. We have seen this explicitly
in the case of the
$D=10$ type II D-branes $(\epsilon =1)$ and the type II* E-branes
$(\epsilon =-1)$ in section 3,
but this applies generally, as can be seen from a simple investigation
of the equations of motion.

Next, suppose that for a given spacetime signature $(S,T)$  and $\epsilon$
there is an  $(s,t)$-brane solution with
world-volume signature $(s,t)$.
Then there will also be $(s',t')$-brane solutions of the same theory
(with the same $(S,T)$  and
$\epsilon$) for all
$(s',t')$ with
 $s'=s+2m$, $t'=t-2m$ for   integers $m$ (positive or negative) such that
$s'\le S$ and $t'\le T$. Again, this is easily seen by considering
the equations of motion.

Finally, for general $(S,T)$,  if $\epsilon =1$
there are electric $(s,t)$ branes   for all odd $t$ with $s+t=n$ and
$0\le s\le S,0\le t\le T$,
(including an electric
$(n-1,1)$-brane   if $n-1\le S$), while if $\epsilon =-1$ there are electric
$(s,t)$ branes
for all
even $t$ with $s+t=n$ and
$0\le s\le S,0\le t\le T$ (including an electric
$(n,0)$-brane  if $n\le S$).

Similarly, starting from the dual action \mgendu, we find that
if $\ti \epsilon =1$
there are magnetic $(s,t)$ branes   for all odd $t$ with $s+t=\ti n$ and
$0\le s\le S,0\le t\le T$,
  while if $\ti \epsilon =-1$ there are magnetic $(s,t)$ branes
for all
even $t$ with $s+t=\ti n$ and
$0\le s\le S,0\le t\le T$.

This agrees   with the branes in 11 dimensions displayed in table 1 and
  the   brane
spectra for the IIA and IIB theories given in tables 2 and 3.

\newsec{Dualities, Reductions and Oxidations}

The various M, IIA and IIB theories are linked by a web of
dualities given in \christwo, and
these should map the branes in the various theories into one another,
giving a check on our
results.
The 11-dimensional theory with signature $(S,T)$ gives the IIA theory with
signature
$(S-1,T)$ on
dimensional reduction on a spatial circle,
and the IIA theory with signature $(S,T-1)$ on
dimensional reduction on a timelike circle.
Then the M-branes   give rise to the various branes of the IIA theory by
simple and double dimensional reductions.

In the usual case of the spatial
reduction of 10+1 dimensional M-theory to the IIA string theory in
9+1 dimensions, the solutions  of M-theory preserving half the supersymmetry
each gives two solutions of the IIA theory preserving half the supersymmetry,
one via simple
dimensional reduction, the other via double dimensional reduction.
The reductions are illustrated in fig.1:

\vskip 0.5cm

{\vbox{
\let\picnaturalsize=N
\def\picsize{6.0in}
\def\picfilename{ramfig3.eps}
\ifx\nopictures Y\else{\ifx\epsfloaded Y\else\input epsf \fi
\let\epsfloaded=Y
\centerline{\ifx\picnaturalsize N\epsfxsize \picsize\fi
\epsfbox{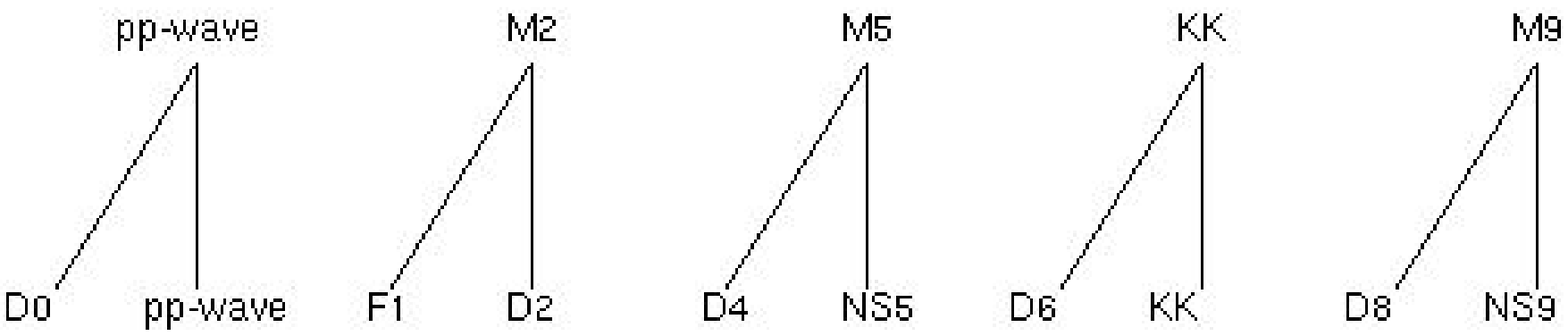}}}\fi
\smallskip
 \leftskip 2pc \rightskip 2pc
 \noindent{
 \ninepoint\sl \baselineskip=8pt
 {\bf Figure 1}
 Spacelike reductions of M  solutions.}
}}
\medskip


It has been   conjectured  \refs{\GravDu,\bergnin}\ 
that there should be some  \lq M9-brane' giving both the D8-brane and
the NS 9-brane \GravDu\ of the IIA theory,  and this has been included.

If instead  M-theory is compactified on a timelike circle to give the  IIA
string theory in
Euclidean dimensions, the reduction of the branes is as in fig. 2.

\vskip 0.5cm

{\vbox{
\let\picnaturalsize=N
\def\picsize{6.0in}
\def\picfilename{ramfig4.eps}
\ifx\nopictures Y\else{\ifx\epsfloaded Y\else\input epsf \fi
\let\epsfloaded=Y
\centerline{\ifx\picnaturalsize N\epsfxsize \picsize\fi
\epsfbox{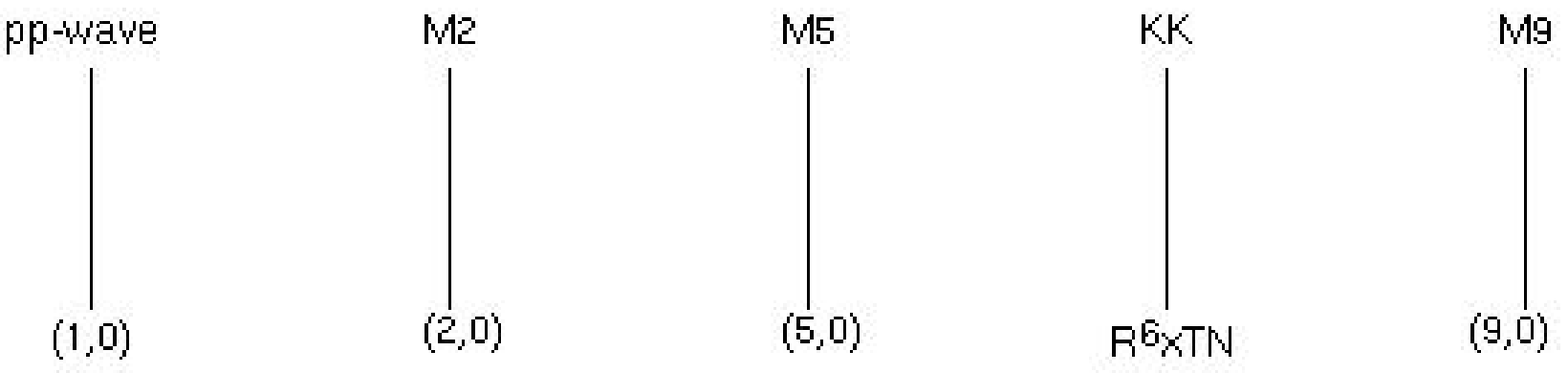}}}\fi
\smallskip
 \leftskip 2pc \rightskip 2pc
 \noindent{
 \ninepoint\sl \baselineskip=8pt
 {\bf Figure 1}
 Timelike reductions of M  solutions.}
}}
\medskip

Note that the world-volume time of a $(p,1)$-brane of M-theory must be wrapped
around the compact
time
of the target space, so that only a double dimensional reduction is possible,
to give a
$(p,0)$-brane.
In this way, all the branes of the IIA$_{10,0}$ theory listed in table 2 are
obtained, along with
a  (9,0)-brane, which was to be expected  by T-duality. Indeed, the
IIA$_{10,0}$ theory is T-dual
to the IIB$'_{9,1}$ string theory, and the T-duality takes the (8,0)-brane
(E8-brane) of the
IIB$'_{9,1}$ string theory to a (9,0) brane of the IIA$_{10,0}$.

Similarly,  the solutions of the M* theory in signature (9,2)
considered in section 5 give
solutions of the IIA*$_{9+1}$ theory via a timelike
reduction and the IIA$_{8+2}$ theory via a spacelike reduction.
Consider first timelike reductions.
{}From the pp-wave of M*, one obtains in IIA* either the ten-dimensional
pp-wave or the E1-brane. From the $(1,2)$-brane one obtains the NS 
(fundamental)
string. {}From the $(3,0)$-brane
we obtain the E3-brane. From the M* fivebrane,
we obtain either the E5-brane or the NS (solitonic) fivebrane.
Finally,  we obtain from
$\R^{5,2} \times TN$ the solution $\R^{5,1} \times TN$,
and from $\R^7\times  TN^{2,2}$ the E7-brane. (Reducing
$\R^7\times  -TN^{2,2}$ on one of the two transverse times,
after taking a periodic array, gives a non-asymptotically flat
solution with a logarithmic harmonic function.) See figure 3.



{\vbox{
\let\picnaturalsize=N
\def\picsize{6.0in}
\def\picfilename{ramfig1.eps}
\ifx\nopictures Y\else{\ifx\epsfloaded Y\else\input epsf \fi
\let\epsfloaded=Y
\centerline{\ifx\picnaturalsize N\epsfxsize \picsize\fi
\epsfbox{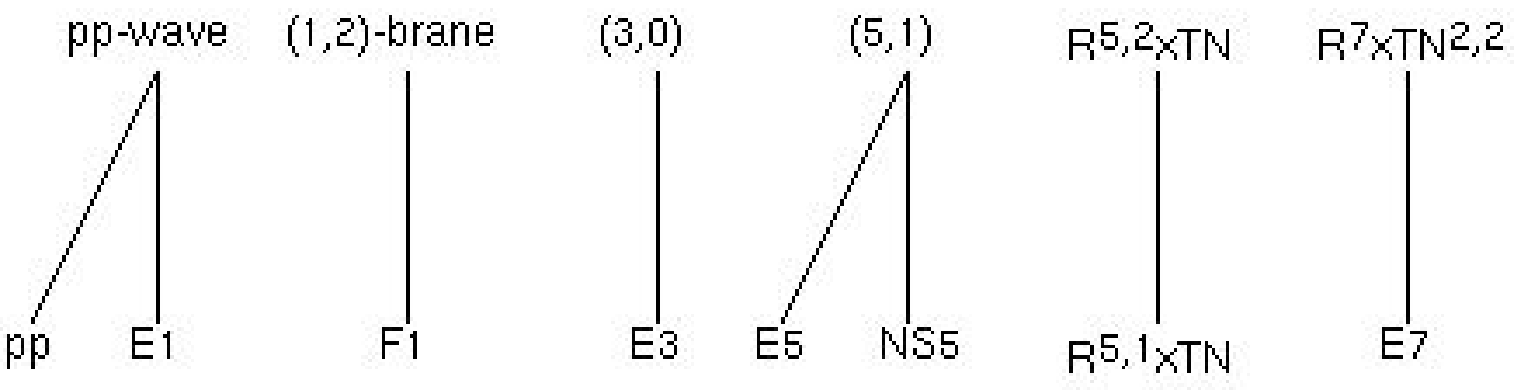}}}\fi
\smallskip
 \leftskip 2pc \rightskip 2pc
 \noindent{
 \ninepoint\sl \baselineskip=8pt
 {\bf Figure 3}
 Timelike reductions of M*  solutions.}
}}
\medskip

Now consider the spacelike reductions. {}From the pp-wave, one obtains
either the pp-wave in IIA$_{8+2}$ or a $(0,1)$ version of the E1-brane
(here $(p,q)$ represents a solution with worldvolume signature $(p,q)$).
{}From the $(1,2)$-brane we obtain 
either a $(1,2)$- or a $(0,2)$-brane. {}From the
$(3,0)$-brane we obtain either a $(3,0)$- or a $(2,0)$-brane, with the other
solutions reducing in a similar fashion. Note that, due to the more
complicated signature of 8+2, there are more solutions in this case than in
9+1. See figure 4.

\vskip 0.5cm


{\vbox{
\let\picnaturalsize=N
\def\picsize{6.0in}
\def\picfilename{ramfig2.eps}
\ifx\nopictures Y\else{\ifx\epsfloaded Y\else\input epsf \fi
\let\epsfloaded=Y
\centerline{\ifx\picnaturalsize N\epsfxsize \picsize\fi
\epsfbox{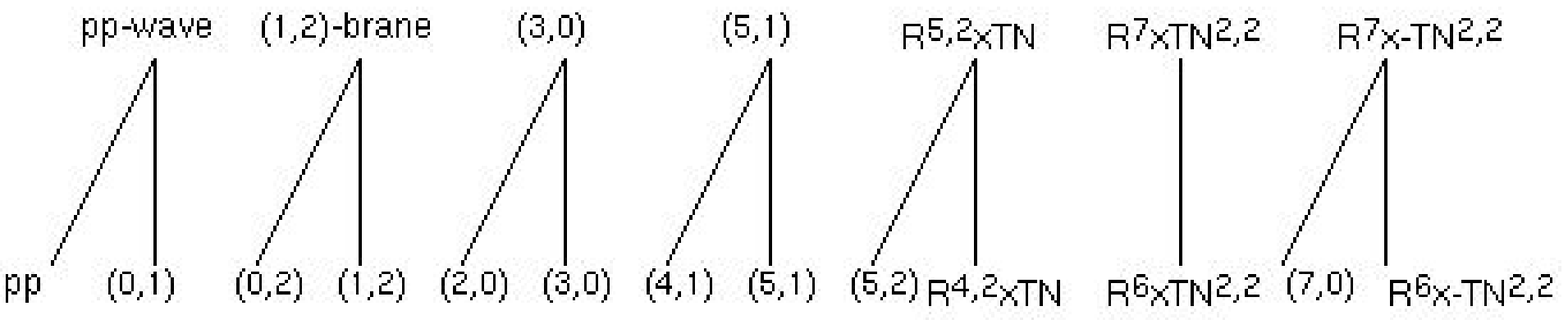}}}\fi
\smallskip
 \leftskip 2pc \rightskip 2pc
 \noindent{
 \ninepoint\sl \baselineskip=8pt
 {\bf Figure 4}
 Spacelike reductions of M*  solutions.}
}}
\medskip

The M$'$ theory in 6+5
dimensions gives the branes of the
usual IIA theory in signature (5,5) on spatial reduction and of the
IIA theory in signature (6,4) on timelike reduction.

A spacelike or timelike T-duality takes a IIA theory to a IIB theory
and vice versa, the precise
relations being given in \christwo. A IIA theory in signature
$(S,T)$ is related via a spatial T-duality (say) to a IIB theory
either in signature
$(S',T')$=$(S,T)$ or in signature $(S',T')=(S-1,T+1)$.
In the former case, the spacelike dimensional reduction of the IIA theory
to $(S-1,T)$ dimensions is
the same as the spacelike reduction of the IIB theory, while in the latter
case it is the same as
the timelike reduction of the IIB theory. This means that a brane of the IIA
theory can be reduced
to a brane of the 9-dimensional supergravity on spatial reduction, and this
can then be oxidised
to a brane of the T-dual IIB theory. Similar results apply for timelike
reductions of the IIA
theory.

For example, the IIA theory in (9,1) dimensions is related by a timelike
T-duality to the IIB*
theory in (9,1) dimensions. Thus the timelike reductions of the D-branes of the
IIA theory
give solutions of a
9-dimensional theory which should give   E-brane   solutions of the
IIB* theory on  timelike oxidation. We now check this.

Following the methods outlined in
Appendix A, suppose we start with the D2-brane solution of IIA
in the canonical frame action
\eqn\actiia{S_{IIA}=\int d^{10} x \sqrt{-g} \left( R -{1\over 8}
(\nabla\Phi)^2 -{1\over 48} e^{\Phi/4} G_4^2\right),}
where all other fields are set to zero. In this frame, the solution is
given by
\eqn\dtwocan{\eqalign{ds^2=&H^{-5/8} (-dt^2 + dx_1^2 + dx_2^2)
+ H^{3/8} (dy_3^2 + \dots  + dy_9^2),\cr
C_{012}=&-H^{-1},\qquad\qquad e^{2\Phi}=H,\cr}}
where $H=1+Q/y^5$ is a harmonic function solving the
Laplace equation in seven (transverse) dimensions
($y^2 =y_3^2 + \dots +y_9^2$).

If we reduce this solution along the $t$ direction following
the first method outlined in Appendix A,
we obtain a Euclideanised string solution
\eqn\eucstring{\eqalign{ds^2=&H^{-5/7} (dx_1^2 + dx_2^2)
+ H^{2/7} (dy_3^2 + \dots  + dy_9^2),\cr
B_{12}=&-H^{-1},\qquad\qquad e^{\phi'}=H\cr}}
to the equations of motion derived from the action
\eqn\acteuc{S_9=\int d^9 x \sqrt{g} \left( R -{1\over 7}
(\nabla\phi')^2 +{1\over 12} e^{4\phi'/7} G_3^2\right).}
Note the change in sign of the last term of the action.

Now if we oxidise vertically along a timelike
direction $\ti t$ following the second method outlined
in Appendix A, we obtain the solution
\eqn\etwo{\eqalign{ds^2=&H^{-3/4} (dx_1^2 + dx_2^2)
+ H^{1/4} (-d\ti t^2+dy_3^2 + \dots  + dy_9^2),\cr
B_{12}=&-H^{-1},\qquad\qquad e^{\phi}=H,\cr}}
to the IIB* action
\eqn\actiibstar{S_{IIB*}=\int d^{10} x \sqrt{-g} \left( R -{1\over 8}
(\nabla\phi)^2 + {1\over 12} e^{\phi/2} G_3^2\right).}
This solution can then immediately be extended to the E2-brane
solution of IIB*
above by allowing the harmonic function to
include $\ti t$-dependence, thus replacing $H$ by \ek{
H'=1+{Q\over (y^2-\ti t^2)^3}.}
Transforming to the sigma-model frame, we recover the solution
\ebr\ for $p=2$.

{}Following this method, it is not difficult to see that the
timelike reduction followed by oxidation procedure transforms the
D$p$-brane into an E$p$-brane,   with $p$ even for IIA and $p$
odd for IIB.

As an example in which the T-duality changes the signature, the  IIA$_{8+2}$
on a spacelike circle
of radius $R$ is T-dual to the IIB$_{7+3}$ theory on a timelike circle of
radius $1/R$.
The spacelike reduction of the
IIA$_{8+2}$ supergravity and the timelike reduction of the IIB$_{7+3}$
supergravity give the same
supergravity theory in $7+2$ dimensions. A spacelike reduction followed
by a timelike oxidation
should then take a brane of the IIA$_{8+2}$  theory to one of the
IIB$_{7+3}$ theory. We now check this.

Suppose we start with the $(3,0)$-brane solution of IIA$_{8+2}$
in the canonical frame action
\eqn\actiiaeighttwo{S_{IIA8+2}=\int d^{10} x \sqrt{g} \left( R -{1\over 8}
(\nabla\Phi)^2 +{1\over 48} e^{\Phi/4} G_4^2\right),}
where all other fields are set to zero. In this frame, the
$(3,0)$ solution is given by
\eqn\dtwocan{\eqalign{ds^2=&H^{-5/8} (dx_1^2 + dx_2^2+dx_3^2)
+ H^{3/8} (-d\ti t_1^2 -d\ti t_2^2 + dy_1^2 + \dots  + dy_5^2),\cr
C_{123}=&-H^{-1},\qquad\qquad e^{2\Phi}=H,\cr}}
where
\ek{
H=1+{Q\over (y^2-\ti t_1^2-\ti t_2^2)^{5/2}}}
is a harmonic function solving the
wave equation in the seven (transverse) dimensions
($y^2 =y_1^2 + \dots +y_5^2$).

If we reduce this solution along the $x_3$ direction following
the first method outlined in \rust\ for spatial reductions,
we obtain a Euclideanised string solution
\eqn\eucstringsevtwo{\eqalign{ds^2=&H^{-5/7} (dx_1^2 + dx_2^2)
+ H^{2/7} (-d\ti t_1^2 -d\ti t_2^2+dy_1^2 + \dots  + dy_5^2),\cr
B_{12}=&-H^{-1},\qquad\qquad e^{\phi'}=H\cr}}
to the action
\eqn\acteucsevtwo{S_{7+2}=\int d^9 x \sqrt{g} \left( R -{1\over 7}
(\nabla\phi')^2 +{1\over 12} e^{4\phi'/7} G_3^2\right).}
Note that there is no change in the sign of the last term of the action.

If we oxidise vertically along a timelike
direction $\ti t_3$, we obtain the solution
\eqn\etwo{\eqalign{ds^2=&H^{-3/4} (dx_1^2 + dx_2^2)
+ H^{1/4} (-d\ti t_1^2 -d\ti t_2^2-d\ti t_3^2+dy_1^2 + \dots  + dy_5^2),\cr
B_{12}=&-H^{-1},\qquad\qquad e^{\phi}=H,\cr}}
to the IIB$_{7+3}$ action
\eqn\actiibstar{S_{IIB7+3}=\int d^{10} x \sqrt{-g} \left( R -{1\over 8}
(\nabla\phi)^2 + {1\over 12} e^{\phi/2} G_3^2\right).}
This solution can then immediately be extended to the $(2,0)$
solution of IIB$_{7+3}$ of section 5
by allowing the solution of the wave equation to
include $\ti t_3$-dependence, thus replacing $H$ by
\ek{
H'=1+{Q\over (y^2-\ti t_1^2-\ti t_2^2-\ti t_3^2)^3}.}
Transforming to the sigma-model frame, we recover the $(2,0)$ solution.

\vskip1truecm

\noindent
{\bf Acknowledgements:}

CMH is supported by an EPSRC Senior Fellowship. RRK is supported
by a PPARC Advanced Fellowship.

\appendix{A}{Timelike Reduction and Oxidation}

Double-dimensional reduction is a procedure relating one class of
$(p+1)$-branes in a $(D+1)$-dimensional theory to another of $p$-branes
in $D$ dimensions. The inverse procedure going from $D$ to $D+1$
dimensions is known as ``oxidation''.
Double-dimensional reduction was originally used to relate the
$D=11$ supermembrane action to that for the Type IIA superstring
in $D=10$ \eleven. The procedure was extended to generate new
solutions in general dimensions in \refs{\ddduff,\stain}.
In \rust, double dimensional reduction on a spatial direction
was extended in a
straightforward manner for non-static $p$-brane solutions
in theories with an arbitrary dilaton coupling constant.
Here we follow \rust\ but reduce on a timelike direction.

The basic approach is to start with a $p+1$-brane solution
in $D+1$ dimensions with arbitrary signature,
and eliminate one of the timelike directions
parallel to the worldvolume of the brane to produce a
Euclideanised $p$-brane in
$D$ dimensions. We begin with the action:
\eqn\rustone{
I={1\over16\pi\hat{G}}\int d^{D+1}\!x\,\sqrt{-\hat{g}}\left[\hat{R}
-{\hg\over2}(\nabla\hat{\phi})^2
\mp {1\over 2(d+2)!}e^{-\ha\hg\hat{\phi}}\hat{F}^2
\right],}
where $\hat{\gamma}=2/(D-1)$.
Here $\hat{F}$ is a $(d+2)$-form field strength defined in terms
of a $(d+1)$-form potential $\hat{A}$
--- \ie $\hat{F}=d\hat{A}$. In the following, it will be useful to
define $\td=D-d-2$. Since we wish to allow
the consideration of anisotropic branes, we will not
restrict our discussion to the choices $p=d-1$ or $\td-2$
for the $(p+1)$-branes in $D+1$ dimensions.

We require that in the solutions all of the fields are independent
of (at least) one of the temporal coordinates, denoted $t$, which
runs parallel to the $(p+1)$-brane. This coordinate will be the
direction which is removed in the double-dimensional reduction.
Further we will require that in the $D+1$ dimensional solution
all of the
nonvanishing components of the $\hat{A}$ potential carry a $t$
index. In the dimensionally reduced
theory, we will have a $d$-form potential $A$ with
$A_{\mu\cdots\nu}=\hat{A}_{\mu\cdots\nu t}$
and a corresponding $(d+1)$-form field strength $F=dA$.
We also make a Kaluza-Klein reduction of the metric
which is required to have the form
\eqn\rusttwo{
\hat{g}_{\mu\nu}=\pmatrix{\bar{g}_{\mu\nu}&0\cr
                          0&-\exp[2\rho]\cr},}
where $-\exp[2\rho]$ is the component $\hat{g}_{tt}$.
One finds that $\hat{R}(\hat{g})=\bar{R}(\bar{g})-2{\bar{\nabla}}^2\rho
-2(\bar{\nabla}\rho)^2$. Thus the gravity part of the action becomes
\eqn\rustthree{
\int d^{D+1}\!x\, \sqrt{-\hat{g}}\,\hat{R}= L
\int d^{D}\!x\, \sqrt{-\bar{g}}\,e^\rho\bar{R},}
up to surface terms,
where $L$ is the volume of the Euclideanised $t$ dimension.
To remove the exponential factor in this action, we make a further
conformal transformation:
$\bar{g}_{\mu\nu}=\exp[-2\rho/(D-2)]g_{\mu\nu}$.
The Ricci scalar then becomes
\eqn\rustfour{
\bar{R}=\exp\left[{2\rho\over D-2}\right]\left(R
+{D-1\over D-2}\nabla^2\rho-{D-1\over D-2}(\nabla\rho)^2\right),}
and now the full dimensionally reduced action is
\eqn\rustfive{
I={L\over 16\pi\hat{G}}\int d^Dx \sqrt{-g}\left[R-{D-1\over
D-2}(\nabla\rho)^2-{\hg\over 2}(\nabla\hat{\phi})^2
\pm {1\over 2(d+1)!}e^{-\ha\hg\hat{\phi}}e^{-2\td\rho/(D-2)}F^2\right].}
Note the sign change in the $F^2$ term.

{}From the equations of motion, we find that
$\nabla^2(\hat{\phi}-(D-1)\ha/\td\ \rho)=0$, and so this linear
combination represents a free scalar field which we have the liberty
to set to zero. Hence setting
$\rho=[\td/((D-1)\ha)]\hat{\phi}$ in the above action, we find
the kinetic term of the remaining scalar has an unconventional
normalization. Thus we scale this field to define the canonical
dilaton $\phi$ of the dimensionally reduced theory
\eqn\rustsix{
{\gamma\over2}(\nabla\phi)^2=
\left({\hg\over2}+{\td^2\over(D-2)(D-1)\ha^2}\right)(\nabla\hat{\phi})^2,}
where $\gamma=2/(D-2)$.
With these choices the final action becomes
\eqn\rustsixa{
I={1\over16\pi G}\int
d^D\!x\,\sqrt{-g}\left[R-{\gamma\over2}(\nabla\phi)^2 \pm {1\over2(d+1)!}
e^{-a\gamma\phi}F^2\right],}
where $a^2=[(D-2)/(D-1)]\ha^2+\td^2/(D-1)$
and $G=\hat{G}/L$.

We can reverse this reduction process to construct an
oxidation prescription as follows: In the $D$-dimensional
theory described by the final action \rustsixa, we begin
with a solution with a given field configuration
$g_{\mu\nu},\ \ \phi,\ \ A$. We add an extra timelike dimension $t$ to
construct a solution for the initial
$(D+1)$-dimensional action \rustone\ as
\eqn\rustseven{
\hat{A}_{\mu\cdots\nu t}=A_{\mu\cdots\nu}, \qquad\qquad
{\hphi/\ha}={\phi/a},}
\eqn\rusteight{
\hat{g}_{\mu\nu}=\pmatrix{\exp\left[-{2\td\over(D-2)(D-1)}{\phi
\over a}\right]
g_{\mu\nu}&0\cr
0&-\exp\left[{2\td\over(D-1)}{\phi\over a}\right]\cr},}
with $\ha^2=[(D-1)/(D-2))]a^2 - \td^2/(D-2)$.

Applying Poincar\'e duality to the field strengths in the
above construction, one arrives at a distinct reduction/oxidation
scheme. First one replaces the $\hat{F}$ in the action
\rustone\ by the dual $(\td+1)$-form
field strength $\hat{H}=e^{-\ha\hg\hat{\phi}}\hat{\tF}$,
which then satisfies $d\hat{H}=0=d(e^{\ha\hg\hat{\phi}}\hat{\tH})$.
Hence the new field strength can be defined in terms
a $\td$-form potential $\hat{B}$, \ie $\hat{H}=d\hat{B}$,
and the new equations of motion will arise from the following
action
\eqn\rustnine{
I={1\over16\pi\hat{G}}\int d^{D+1}\!x\,\sqrt{-\hat{g}}\left[\hat{R}
-{\hg\over2}(\nabla\hat{\phi})^2
\mp {1\over 2(\td+1)!} e^{\ha\hg\hat{\phi}}\hat{H}^2
\right]\ .}
In the dimensional reduction with the dual field,
any components of $\hat{B}$ carrying a $t$ index will vanish,
\ie, $\hat{B}_{\mu\cdots\nu t}=0$. Further
in the reduced theory, we have a $\td$-form potential $B$ given by
$B_{\mu\cdots\nu}=\hat{B}_{\mu\cdots\nu}$,
and a corresponding $(\td+1)$-form field strength $H=dB$.
The remainder of the construction is unchanged.
The essential point though is that one arrives at a second
independent reduction/oxidation procedure in which the form
potential is completely unchanged. In this case, however,
the sign of the $F^2$ term is unchanged.

Expressed in terms of the action \rustsixa, the second
oxidation prescription is as follows: We begin in the $D$-dimensional
theory described by the final action \rustsixa\ with a
solution given by the field configuration
$g_{\mu\nu},\ \ \phi,\ \ A.$
We add an extra dimension $t$ to
construct a solution for a $(D+1)$-dimensional theory with the action
\eqn\rustten{
I={1\over16\pi\hat{G}}\int d^{D+1}\!x\,\sqrt{-\hat{g}}\left[\hat{R}
-{\hg\over2}(\nabla\hat{\phi})^2
\pm {1\over 2(d+1)!}e^{-\ha\hg\hat{\phi}}{F}^2
\right],}
where $\ha^2=[(D-1)/(D-2))]a^2 - d^2/(D-2)$.
Also ${F}=d{A}$ is the same $(d+1)$-form field strength that
appears in the original action. The field configuration of
the $(p+1)$-brane solution leaves the nonvanishing
components of the $d$-form potential $A$ unchanged, and has
\eqn\rusteleven{
{\hphi/\ha}={\phi/a},}
\eqn\rusttwelve{
\hat{g}_{\mu\nu}=\pmatrix{\exp\left[{2d\over(D-2)(D-1)}{\phi
\over a}\right]g_{\mu\nu}&0\cr
0&-\exp\left[-{2d\over(D-1)}{\phi\over a}\right]\cr}.}
Again, the sign of the $F^2$ term is unchanged.

Thus beginning with $p$-brane solutions of one theory,
we have two independent oxidation procedures.
The first prescription
extends the form degree of the potential $A$, as well as the dimension
of the brane
and the spacetime, while the second
leaves the degree of the potential unchanged. Of course, the two
oxidation procedures lead to different $(D+1)$-dimensional
actions.

\appendix{B}{Curvature Components for p-brane-Type Ansatzes}

The solution of the equations of motion for theories in different
signatures proceeds along the lines of the $(p,1)$-brane solutions
shown in \prep. The Einstein equations are typically the most
complicated to write down and solve, involving various
components of the Ricci tensor. For reference, we write down
these components for $(p,q)$-brane type ansatzes.

In a $D$-dimensional spacetime signature $(S,T)$, where $S+T=D$, 
consider the following ansatz for the metric:
\eqn\bone{ds^2=e^{2A} \eta_{ab} dX^a dX^b +
e^{2B} \ti \eta_{mn} dY^m dY^n,}
where $\eta_{ab}$ and $\ti \eta_{mn}$ are flat metrics of 
signature $(s,t)$ and $(S-s,T-t)$, respectively.
$a,b$ run over $d=s+t$ indices while $m,n$ run over $S+T-s-t=D-d=\tilde d +2$
indices.
$A$ and $B$ are functions of the $Y^m$ coordinates only.
A straightforward computation leads to the following results for
the Ricci tensor and scalar:
\eqn\btwo{R_{ab}=-\eta_{ab} e^{2(A-B)}
\left( \nabla^2 A + d (\pa A)^2
+ \tilde d \pa A \cdot \pa B\right),}
\eqn\bthree{\eqalign{R_{mn}=& - d \pa_m \pa_n A -\tilde d \pa_m \pa_n B
-\ti \eta_{mn} \nabla^2 B -d \pa_m A \pa_n A + \tilde d \pa_m B
\pa_n B \cr
& + d(\pa_m A \pa_n B + \pa_n A \pa_m B) - d \ti \eta_{mn} \pa A \cdot \pa B
- \tilde d \ti \eta_{mn} (\pa B)^2 \cr}}
and
\eqn\bfour{R=e^{-2B}\left( -2d \nabla^2 A -2(\tilde d + 1) \nabla^2 B
-d(d+1) (\pa A)^2 -\tilde d (\tilde d + 1) (\pa B)^2 -2d \tilde d \pa A
\cdot \pa B
\right),}
where $\nabla^2\equiv \ti \eta^{kl}\pa_k \pa_l$ and 
$V \cdot W \equiv \ti \eta^{kl} V_k W_l$.

\listrefs
\end